\documentclass[10pt,journal]{IEEEtran}
\makeatletter
\def\ps@headings{%
\def\@oddhead{\mbox{}\scriptsize\rightmark \hfil \thepage}%
\def\@evenhead{\scriptsize\thepage \hfil \leftmark\mbox{}}%
\def\@oddfoot{}%
\def\@evenfoot{}}
\makeatother
\usepackage{amssymb}
\usepackage{amsfonts,amsmath}
\usepackage[ruled]{algorithm}
\usepackage{algpseudocode}
\usepackage{algorithmicx}
\usepackage{graphicx,epsfig,subfigure}
\usepackage{subfig}

\usepackage{color}

\usepackage{cite}

\newcommand {\mymarginpar}[1]{\marginpar{#1}}
\renewcommand {\marginpar}[1]{}

\def\_{\rule{.3em}{.15ex}}      

\newcommand{\ls}[1]
   {\dimen0=\fontdimen6\the\font
    \lineskip=#1\dimen0
    \advance\lineskip.5\fontdimen5\the\font
    \advance\lineskip-\dimen0
    \lineskiplimit=.9\lineskip
    \baselineskip=\lineskip
    \advance\baselineskip\dimen0
    \normallineskip\lineskip
    \normallineskiplimit\lineskiplimit
    \normalbaselineskip\baselineskip
    \ignorespaces
   }

\newcommand {\bearn}{\begin{eqnarray*}}
\newcommand {\eearn}{\end{eqnarray*}}
\newcommand {\barr}{\begin{array}}
\newcommand {\earr}{\end{array}}

\newcommand {\N}{{\cal N}}

\newtheorem{definition}{Definition}
\newtheorem{property}[definition]{Property}
\newtheorem{proposition}[definition]{Proposition}
\newtheorem{lemma}[definition]{Lemma}
\newtheorem{theorem}[definition]{Theorem}
\newtheorem{corollary}[definition]{Corollary}
\newtheorem{example}[definition]{Example}
\newtheorem{remark}[definition]{Remark}

\newcommand {\benum} {\begin{enumerate}}
\newcommand {\eenum} {\end{enumerate}}

\newcommand {\bdesc} {\begin{description}}
\newcommand {\edesc} {\end{description}}

\newcommand {\bfig}[2] {\begin{figure}
  \centering
  \includegraphics[width=#2]{#1}}
\newcommand {\brotatefig}[2] {\begin{figure}[htbp]
                        \centerline {
                         \epsfig{figure={#1},clip=,angle=-90,width={#2}}}}
\newcommand {\bfigfirst}[2] {\begin{figure}[h]
                        \centerline {
                        \setlength{\epsfxsize}{#2}
                        \epsffile{#1}}}
\newcommand {\efig}[2]{ \caption{#2}
                        \label{fig:#1}
                        \end{figure}
                        \mymarginpar{fig:#1}}
\newcommand {\erotatefig}[2]{ \caption{#2}
                        \label{fig:#1}
                        \end{figure}
                        \mymarginpar{fig:#1}}

\newcommand {\btab}[1]{
                       \begin{table}
                       \centering
                       \begin{tabular}{#1}}
\newcommand {\etab}[3] {
                       \end{tabular}
                       \caption[#3]{#2}
                       \label{tab:#1}
                       \end{table}
                       \mymarginpar{tab:#1}
                       \vspace{.1in}}

\newcommand {\btabular}[1]{\begin{center}
                       \begin{tabular}{#1}}
\newcommand {\etabular}{\end{tabular}
                       \end{center}}

\newcommand {\bdefin}[1]{\begin{definition}
                      \mymarginpar{def:#1}
                      \label{def:#1} }
\newcommand {\edefin}       {\end{definition}}
\newcommand {\rdef}[1]{Definition \ref{def:#1}}

\newcommand {\bpro}[1]{\begin{property}
                      \mymarginpar{pro:#1}
                      \label{pro:#1} }
\newcommand {\epro}   {\end{property}}

\newcommand {\bprop}[1]{\begin{proposition}
                      \mymarginpar{prop:#1}
                      \label{prop:#1} }
\newcommand {\eprop}       {\end{proposition}}
\newcommand {\rprop}[1]{Proposition \ref{prop:#1}}

\newcommand {\blem}[1]{\begin{lemma}
                      \mymarginpar{lem:#1}
                      \label{lem:#1} }
\newcommand {\elem}   {\end{lemma}}

\newcommand {\bthe}[1]{\begin{theorem}
                      \mymarginpar{the:#1}
                      \label{the:#1} }
\newcommand {\ethe}   {\end{theorem}}
\newcommand {\rthe}[1]{Theorem \ref{the:#1}}

\newcommand {\bproof}{\noindent {\bf Proof.} \ }
\newcommand {\eproof} {\hfill \squares \\ \vspace{.3cm}}
\newcommand {\bcor}[1]{\begin{corollary}
                      \mymarginpar{cor:#1}
                      \label{cor:#1} }
\newcommand {\ecor}   {\end{corollary}}

\newcommand {\bax}[1]{\begin{axiom}
                      \mymarginpar{ax:#1}
                      \label{ax:#1} }
\newcommand {\eax}       {\vspace{-.1in} \end{axiom}}

\newcommand {\bex}[2]{\vspace{.1in}
                      \begin{example}
                      \mymarginpar{ex:#1}
                       {\bf #2}
                      \label{ex:#1} }
\newcommand {\eex}       {\end{example} \vspace{.3cm} }

\newcommand {\brem}[1]{\begin{remark}
                      \mymarginpar{rem:#1}
                      \label{rem:#1} \em }
\newcommand {\erem}   {\end{remark}}

\newcommand {\beq}[1]{\mymarginpar{eq:#1}
                      \begin{equation}
                      \label{eq:#1} }

\newcommand {\beqno}[1]{\mymarginpar{eq:#1}
                      \begin{eqnarray}
                      \nonumber}

\newcommand {\eeq}       {\end{equation}}
\newcommand {\eeqno}       { && \end{eqnarray}}
\newcommand {\req}[1]{(\ref{eq:#1})}

\newcommand {\bear}[1]{\mymarginpar{eq:#1}
                       \begin{eqnarray}
                       \label{eq:#1} }

\newcommand {\bearno}[1]{\mymarginpar{eq:#1}
                       \begin{eqnarray}
                       \nonumber}

\newcommand {\eear}{\end{eqnarray}}
\newcommand {\eearno}{\end{eqnarray}}
\newcommand {\bsel}{\left \{ \begin{array}{cl}}
\newcommand {\esel}{\end{array} \right.}

\newcommand {\bmat}[1]{\left [ \begin{array}{#1}}
\newcommand {\emat}{\end{array} \right ]}
\newcommand {\bsec}[2]{\mymarginpar{sec:#2}
                       \section{#1}
                       \label{sec:#2} }

\newcommand {\rsec}[1]{Section \ref{sec:#1}}

\newcommand {\bsubsec}[2]{\mymarginpar{sec:#2}
                       \subsection{#1}
                       \label{sec:#2} }

\newcommand {\bsubsubsec}[2]{\mymarginpar{sec:#2}
                       \subsubsection{#1}
                       \label{sec:#2} }

\def\R{I\kern-0.30em R}
\def\N{I\kern-0.30em N}
\def\P{I\kern-0.30em P}
\newcommand\squares{\vrule height6pt width7pt depth1pt}

\def\pr{{\bf\sf P}}

\newcommand{\peri}{p}
\newcommand{\rr}[1]{{\color{red} #1}}
\newcommand{\rb}[1]{{\color{blue} #1}}
\newcommand{\rg}[1]{{\color{green} #1}}

\begin{document}

\title{On the Theoretical Gap of Channel Hopping Sequences with Maximum Rendezvous Diversity in the Multichannel Rendezvous Problem}

\author{Cheng-Shang Chang, ~\IEEEmembership{Fellow,~IEEE}, Jang-Ping Sheu,~\IEEEmembership{Fellow,~IEEE}, and Yi-Jheng Lin\\
Institute of Communications Engineering\\
National Tsing Hua University \\
Hsinchu 30013, Taiwan, R.O.C. \\
Email:  cschang@ee.nthu.edu.tw;  sheujp@cs.nthu.edu.tw; s107064901@m107.nthu.edu.tw}

\maketitle

\begin{abstract}
In the literature, there are several well-known periodic channel hopping (CH) sequences that can achieve maximum rendezvous diversity in  a cognitive radio network (CRN). For a CRN with $N$ channels,
it is known that the period of such a CH sequence is at least $N^2$. The asymptotic approximation ratio, defined as the ratio of the period of a CH sequence to the lower bound $N^2$ when $N \to \infty$, is still
2.5 for the best known CH sequence in the literature.
An open question in the multichannel rendezvous problem  is whether it is possible to construct a periodic CH sequence that has the asymptotic approximation ratio 1. In this paper, we tighten the theoretical gap
by proposing CH sequences, called IDEAL-CH, that have the asymptotic approximation ratio 2.

For a weaker requirement that only needs the two users to rendezvous on {\em one} commonly available channel in a period,
 we propose channel hopping sequences, called ORTHO-CH, with period  $(2\peri +1)\peri$,
where $\peri$ is the smallest prime not less than $N$.
\end{abstract}

\begin{IEEEkeywords}
multichannel rendezvous, worst case analysis.
\end{IEEEkeywords}



\bsec{Introduction}{introduction}

The multichannel rendezvous problem that asks two users to find each other by hopping over their available channels is a fundamental problem in cognitive radio networks (CRNs) and has received a lot of attention lately (see, e.g.,  the excellent book \cite{Book} and references therein).
 In this paper, we tighten a theoretical gap on the minimum period of the periodic channel hopping (CH) sequences that achieve maximum rendezvous diversity.
A channel is called a {\em rendezvous} channel of a periodic CH sequence  if  two asynchronous users (with any arbitrary starting times of their CH sequences) rendezvous on that channel within the period of the sequence.
A periodic CH sequence is said to achieve maximum rendezvous diversity for a CRN with $N$ channels if all the $N$ channels are rendezvous channels.
In the asymmetric setting, it was shown in Theorem 1 of \cite{Bian2013} that there do not exist deterministic periodic CH sequences that can achieve maximum rendezvous diversity with periods less than or equal to $N^2$.
For the symmetric setting,
the negative result of Theorem 1 of \cite{Bian2013} is further extended in Theorem 3 of \cite{DRDS13}. It was shown that the length of the period $\peri$ satisfies the following lower bound:
\begin{eqnarray*}
\peri \ge \left\{\begin{array}{ll}
              N^2+N &\mbox{if $N\le 2$}\\
              N^2+N+1 &\mbox{if $N \ge 3$ and $N$ is a prime power}\\
              N^2+2N & \mbox{otherwise}
              \end{array}
              \right..
\end{eqnarray*}
The lower bound is not always tight. Via extensive computer enumeration, it was shown in \cite{DRDS13} that the lower bound
is tight when $N = 1, 2, 5, 6$. It is also tight for $N=8$ by an explicit CH sequence in \cite{Hou2011}.
In the literature, there are various periodic CH sequences that can achieve maximum rendezvous diversity, see, e.g.,
CRSEQ \cite{CRSEQ}, JS \cite{JS2011}, DRDS \cite{DRDS13},  T-CH \cite{Matrix2015}, and DSCR \cite{DSCR2016}. In particular, T-CH \cite{Matrix2015} and DSCR \cite{DSCR2016}
have the shortest period $2N^2 +N\lfloor N/2 \rfloor$ when $N$ is a prime.
These CH sequences are called {\em nearly optimal} CH sequences as their periods are $O(N^2)$, which is comparable to the lower bound $N^2$.
However, the asymptotic approximation ratio, defined as the ratio of the period to the lower bound $N^2$ when $N \to \infty$, is
still 2.5 for T-CH and DSCR, 3 for CRSEQ and DRDS.
One of the open questions in the multichannel rendezvous problem is whether it is possible to construct a periodic CH sequence that has the asymptotic approximation ratio 1.
The main objective of this paper is to further tighten the theoretical gap by proposing CH sequences, called IDEAL-CH, that have the asymptotic approximation ratio 2. To the best of our knowledge, this is the best asymptotic approximation ratio in the literature.

The mathematical tools for the construction of IDEAL-CH are (i) perfect difference sets \cite{Singer1938} and (ii) ideal matrices \cite{Kumar88}.
Intuitively, a perfect  difference set  can be visualized as a one-dimensional (1D) dot pattern that has a dot on the 1D-coordinate of an element. Repeat the dot pattern  infinitely often in the line.
Then for any time shift, exactly one pair of dots will overlap in every period.
On the other hand, an ideal matrix can be viewed as a two-dimensional (2D) version of a perfect difference set.
An  ideal matrix  can be visualized as a 2D dot pattern that has a dot on the 2D-coordinate of an element in the matrix. Repeat the dot pattern  infinitely often in the plane. Then except purely vertical shifts,  exactly one pair of dots will overlap within a  square box for any other two-dimensional shifts. Using different sets for constructing CH sequences is not new (see, e.g., \cite{Hou2011,DRDS13}). However,
it seems that researchers in the field may not be familiar with the concept of ideal matrices. To our surprise, we find out that the constructions of CRSEQ \cite{CRSEQ}, T-CH \cite{Matrix2015}, and DSCR \cite{DSCR2016}, are all based on ideal matrices and they are ``equivalent'' in that sense.
To deal with the problem of purely vertical shifts, CRSEQ, T-CH, and DSCR all add a ``stay'' matrix in front of a ``jump'' matrix constructed from an ideal matrix. The added ``stay'' matrix increases the length of a CH sequence.
To push the asymptotic approximation ratio further down, our idea is to embed different sets into an ideal matrix.
By doing so, we are able to eliminate the need for adding a ``stay'' matrix and thus shorten the length of IDEAL-CH.

CRSEQ, JS, DRDS, T-CH and DSCR, and IDEAL-CH are sequences that can achieve maximum rendezvous diversity within their periods.
A weaker requirement is to ask the two users to rendezvous on {\em one} commonly available channel and measure the maximum time-to-rendezvous (MTTR).
For this, we propose a CH sequence, called ORTHO-CH, which can guarantee  the rendezvous of the two users within a period of the ORTHO-CH sequence.
When the available channels of a user  is a subset of the $N$ channels, the period of our ORTHO-CH sequence is $(2\peri +1)\peri$,
where $\peri$ is the smallest prime not less than $N$.
Thus, ORTHO-CH has the MTTR bound $(2\peri +1)\peri$.
Such a result is comparable to the best algorithms in the literature, e.g., FRCH \cite{ChangGY13} with the MTTR bound $(2N+1)N$ for $N \ne ((5+2\alpha)*r-1)/2$ for all integer $\alpha \ge 0$ and odd integer $r \ge 3$, and SRR　\cite{Localimprove2018} with the MTTR bound $2\peri^2 +2\peri$.

The paper is organized as follows: In \rsec{review}, we provide a brief review of the multichannel rendezvous problem, including the classification of the problem in \rsec{classification}, the formulation of the problem  and summaries of known results  in \rsec{problem}. In \rsec{ICH}, we propose the IDEAL-CH sequences that have the asymptotic approximation ratio 2.
By extending the mathematical theories for IDEAL-CH,  we  propose in \rsec{orthoCH} the ORTHO-CH sequences that have the MTTR bound $\peri (2\peri +1)$, where $\peri$ is the smallest prime not less than the total number of channels.
The paper is  concluded in \rsec{con}.

\bsec{The multichannel rendezvous problem}{review}

In this section, we provide a brief review of the multichannel rendezvous problem (MRP). 

\bsubsec{Classification of the problem}{classification}

\index{Multichannel rendezvous problem (MRP)!classification}

As mentioned in the Introduction, the multichannel rendezvous  problem that asks two {\em users} to find each other by hopping over  a set of possible {\em channels (discrete locations)} with respect to {\em time}. In view of this, there are three key elements in the multichannel rendezvous problem: (i) users, (ii) time, and (iii) channels.
Based on the assumptions on {\em users, time, and channels}, CH schemes can be classified into various settings.
To compare the level of difficulty between two settings $A$ and $B$, we use the partial ordering $A \prec B$ when the assumption in setting $A$ is {\em stronger} than that in $B$. Thus, the CH sequences constructed by using a {\em weaker} assumption in setting $B$ are also applicable in setting $A$.

\noindent 1) {\bf users:}

\index{Symmetric (sym)}
\index{Asymmetric (asym)}
 There are three commonly used settings for users: (i) the symmetric setting ({\em sym} for short), (ii) the ID setting ({\em ID} for short),
 and (iii) the asymmetric setting ({\em asym} for short).
 In the {\em symmetric} setting, users are {\em indistinguishable} and thus follow the same algorithm to generate their CH sequences.
 On the other hand, users are {\em distinguishable} by their unique identifiers (ID) in the ID setting. For instance, a device in a CRN may be equipped with a unique 48-bit medium access control (MAC) address. The asymmetric setting is a special case of the unique ID setting when these two users can be distinguished by  one bit ID, e.g., user 1 is assigned with ID 0 and user 2 is assigned with ID 1.
  In the asymmetric setting, the two users can be assigned two different roles so that they can follow two different algorithms to generate their CH sequences. In the literature, these CH algorithms are called {\em role-based CH algorithms}.  For instance, a user can be assigned the role of a sender or the role of a receiver.  The receiver can stay on the same channel while the sender cycles through all the available channels.
   Since users follow different algorithms, the time-to-rendezvous can be greatly reduced by using role-based algorithms.
 In the general ID setting,  a common approach is to map an ID into an $M$-bit binary vector and partition the time into intervals with $M$ time slots. Then ask each user to play the role in  the $\ell^{th}$ time slot in an interval according to the $\ell^{th}$ bit in the mapped binary vector.
 However, using IDs to generate CH sequences might be vulnerable to attacks from adversaries. As such, it is preferable to remain anonymous in practice.

 In the symmetric setting, the two users are indistinguishable. The key in the symmetric setting is to {\em break symmetry}. One way to break symmetry is to select a channel from the available channel set of a user and use that as the ID of a user. One problem for that is when the two users select the same channel and thus have the same ID.
  In the level of difficulty of the three settings for users,
 $$asym \prec ID \prec sym.$$


\noindent 2) {\bf Time:}

\index{Synchronous (sync)}
\index{Asynchronous (async)}
For the multichannel rendezvous problem, we only consider the discrete-time setting, where time is indexed from $t=0,1,2,\ldots$.
There are two settings for time: (i) the synchronous  setting ({\em sync} for short) and (ii) the asynchronous  setting ({\em async} for short).
In the {\em synchronous}  setting, the clocks (i.e., the indices of
time slots) of both users are assumed to be synchronized to the  global clock and thus the time indices of these two users are  the same.
When the clocks of the two users are synchronized, both users can start their CH sequences simultaneously to speed up the rendezvous process. 
On the other hand, in the {\em asynchronous}  setting, the clocks of both users may not be synchronized to the  global clock
and thus  the time indices of these two users might be different.
In a distributed environment, it might not be practical to assume that the clocks of two users are synchronized as they have not rendezvoused yet. Without clock synchronization, guaranteed rendezvous is much more difficult.
In the level of difficulty of the two settings for time,
 $$sync \prec async.$$

\noindent 3) {\bf Available channels (search space):}

\index{Available channels}
\index{Homogeneous (homo)}
\index{Heterogeneous (hetero)}

For the multichannel rendezvous problem, we only consider distinct channels (discrete locations in \cite{AW90}) as the search space. These $N$ channels are indexed from $0,1,\ldots, N-1$.
The available channel set of a user is a subset of these $N$ channels.
There are two settings for available channels: (i) the homogeneous  setting ({\em homo} for short) and (ii) the heterogeneous setting ({\em hetero} for short).
In the {\em homogeneous} setting,  the available channel sets of the two users are assumed to be the same. On the other hand, in the heterogeneous setting, the available channel sets of the two users might be different. In a CRN, two users that are close to each other are likely to have the same available channel sets. Due to the limitation of the coverage area of a user, two users tend to have different available channel sets if they are far apart. Rendezvous in a homogeneous environment is in general much easier than that in a heterogeneous environment.
In the level of difficulty of the two settings for available channels,
$$homo \prec hetero.$$

\noindent 4) {\bf Labels of channels:}
\index{Labels of channels}
\index{Global}
\index{Local}

There are three widely used settings for the labels of channels: (i) the globally labelled setting  ({\em global} for short),
(ii) the locally labelled setting  ({\em local} for short), and (iii) the indistinguishable setting ({\em ind} for short).
In the multichannel rendezvous problem, the $N$ channels are commonly assumed to be {\em globally labelled}, i.e.,
the labels of the channels of the two users are the same.
On the other hand, the users are only allowed to label their available channels by themselves in the
 {\em locally labelled}  setting.  In the locally labelled setting, the labels of channels could be different.
In the book \cite{Book}, the locally labelled setting is referred to as the  {\em oblivious} setting.
The most difficult setting for labels of channels is where users are not allowed to leave any marks for channels (see, e.g., \cite{DFKP06}). In such a setting, these $N$ channels are {\em indistinguishable} and a user even does not know the previous channels on which it hops. Thus, nothing can be learned from a failed attempt to rendezvous in the indistinguishable setting. In the level of difficulty of the three settings for labels of channels,
$$global \prec local \prec ind.$$

Like the notations in queueing theory, a multichannel rendezvous problem (MRP) can be described by a series of abbreviations and slashes such as
$$A/B/C/D,$$ where $A$ is the abbreviation for the setting of {\em users},  $B$ is the abbreviation for the setting of {\em time},
$C$ is the abbreviation for the setting of {\em available channels}, and $D$ is the abbreviation for the setting of {\em labels of channels}.
 For instance, the sym/async/hetero/global MRP
denotes the problem where (i) the two users are symmetric and thus follow the same algorithm, (ii) the clocks of the two users are not synchronized, (iii) the available channel sets of the two users are different, and (iv) the channels are globally labelled.

We note that there are five categories for the classification of the multichannel rendezvous problem in the book \cite{Book}:
$<$ Alg, Time, Port, ID, Label $>$. Here we combine the {\em Alg} (algorithm) category and the {\em ID} category into our {\em user} category.
Also, the symmetric (resp. asymmetric) port setting in \cite{Book} corresponds to the homogeneous (resp. heterogeneous) setting in which the two users have the same (resp. different) available channel sets.
Thus, the  four categories in our classification are basically the same as the five categories
 in \cite{Book}. 

\bsubsec{Mathematical formulation of the  problem}{problem}

\index{Multichannel rendezvous problem (MRP)!formulation}
  To formulate the multichannel rendezvous problem (MRP), let us  consider a CRN with $N$ channels (with $N \ge 2$), indexed from $0$ to $N-1$. There are two (secondary) users who would like to rendezvous on a common unblocked channel by hopping over these  channels with respect to time. We assume that time is slotted (the discrete-time setting) and indexed from $t=0,1,2,\ldots$. The length of a time slot, typically in the order of 10ms, should be long enough for the two users to establish their communication link on a common unblocked channel. In the literature, the slot boundaries of these two users are commonly assumed to be aligned. In the case that the slot boundaries of these two users  are not aligned, one can double  the size of each time slot so that the overlap of two misaligned time slots is not smaller than the original length of a time slot.

  The available channel set for user $i$, $i=1,2$, $${\bf c}_i=\{c_i(0), c_i(1), \ldots, c_i(n_i-1)\},$$
  is a subset of the $N$ channels. Let $n_i=|{\bf c}_i|$ be the number of available channels to user $i$, $i=1,2$.  In the homogeneous setting,
  the available channel set for each user is simply the set of the $N$ channels, i.e.,
  $${\bf c}_1={\bf c}_2=\{0,1,\ldots, N-1\}.$$

  We assume that there is at least one channel that is commonly available to the two users (as otherwise it is impossible for the two users to rendezvous), i.e.,
\beq{avail1111}
{\bf c}_1 \cap {\bf c}_2 \ne \varnothing.
\eeq
  Denote by $X_1(t)$ (resp. $X_2(t)$) the  channel selected by user 1 (resp. user 2)  at time $t$ (of the global clock). 
  Then the time-to-rendezvous (TTR), denoted by  $T$, is the number of time slots (steps) needed for these two users to select a common available channel, i.e.,
\beq{meet1111}
T=\inf\{t \ge 0: X_1(t) = X_2(t) \}+1,
\eeq
where we add 1 in \req{meet1111} as we start from $t=0$.

In addition to the time-to-rendezvous, we are also interested in the time to achieve maximum rendezvous diversity, denoted by  $T^\sharp$, that is defined as the first time that the two users have met each other on  every commonly available channel.
Specifically,
let $T_i$ be the first time that these two users hop on channel $i$ at the same time,
i.e.,
\beq{meet1111i0}
T_i=\inf\{t \ge 0:  X_1(t) =X_2(t)=i\}+1.
\eeq
  Then
\beq{meet1111sharp0}
T^\sharp=\max_{ i \in {\bf c}_1 \cap {\bf c}_2}T_i.
\eeq
Note that
  $T$ can also be presented as follows:
\beq{meet1111min0}
T=\min_{ i \in {\bf c}_1 \cap {\bf c}_2}T_i.
\eeq
Clearly, $T \le T^\sharp$.
We say a CH scheme has a maximum time-to-rendezvous (MTTR)  bound $\gamma$ (for some finite constant $\gamma$) if $T \le \gamma$.
Similarly, a CH scheme has a maximum conditional time-to-rendezvous (MCTTR) bound $\gamma$ if $T^\sharp \le \gamma$.

In the literature, there are three commonly used metrics for evaluating the performance of a CH sequence:
\begin{description}
\item[(i)] expected time-to-rendezvous (ETTR),
\item[(ii)] maximum time-to-rendezvous (MTTR), and
\item[(iii)] maximum conditional time-to-rendezvous (MCTTR).
\end{description}

The simplest way to generate  CH sequences  is  the {\em random} algorithm that  selects a channel uniformly at random in a user's available channel set in every time slot. The random algorithm performs amazingly well in terms of ETTR and its ETTR is quite close to the lower bound in the the asym/async/hetero/local MRP (see, e.g.,  \cite{ToN2017}). As such, it outperforms many CH algorithms proposed in the literature in terms of ETTR. However, the random algorithm does not have bounded MTTR. Thus, for theoretical analysis, researchers in the field focus mostly on MTTR/MCTTR.

In Table \ref{table:known}, we provide a summary for the known results of various rendezvous algorithms in their most difficult settings.

{
\tiny
\begin{table*}
\begin{center}
\caption{Known results of various rendezvous algorithms in their most difficult settings.}
\begin{tabular}{||c||c|c|c|c|c|c||} \hline\hline
& users & time & channels & labels &  MTTR/MCTTR & ETTR \\ \hline
WFM \cite{AW90,ToN2015} & asym & async & homo & local &  $N$   & $\frac{N+1}{2}$ \\ \hline
WFM-MRD \cite{ToN2015} & asym & async & hetero & local &    $N^2$ (MRD)&  \\ \hline
AFCHS \cite{AFCHS2018} & asym & async & hetero & global &   $N^2$ (MRD)&  \\ \hline
coprime MC \cite{Theis2011,ToN2017} & asym & async & hetero & local &  $2(n_1+1)n_2$ (MRD) &  \\ \hline \hline
FOCAL \cite{AG03} & sym & async & homo & global & 1   & 1 \\ \hline
SynMAC \cite{SynMAC} & sym & sync & hetero & global &   $N$ (MRD) &  \\ \hline
M-QCH \cite{Quorum} & sym & sync & hetero & global &   $3N$ (MRD) &  \\ \hline
SSCH \cite{SSCH} & sym & sync & homo & global &  $N+1$   & $\frac{N+1}{2}+\frac{1}{2}-\frac{1}{2N}$ \\ \hline
FPP \cite{MOR2014} & sym & sync & homo & global &  $N+1$   & $\frac{N+1}{2}+\frac{1}{2}-\frac{1}{2N}$  \\ \hline
RRICH \cite{ToN2015} & sym & sync & hetero & global &  $N(N+1)$ (MRD) &  \\ \hline
CACH \cite{ToN2015} & sym & sync & hetero & global &  $N(u+1)$ (MRD) &  \\ \hline \hline
SeqR \cite{Seq2008} & sym & async & homo & global &  $N(N+1)$   &  \\ \hline
DRSEQ \cite{DRSEQ} & sym & async & homo & global &  $2N+1$   & $N-\frac{1}{6}+\frac{2N^2+11N-4}{6N(2N+1)^2}$ \\ \hline \hline
JS \cite{JS2011} & sym & async & hetero & global &    $3N^3+o(N^3)$ (MRD) &  \\ \hline
CRSEQ \cite{CRSEQ} & sym & async & hetero & global &    $P(3P-1)$ (MRD) &  \\ \hline
DRDS \cite{DRDS13} & sym & async & hetero & global &   $3P^2$ (MRD) &  \\ \hline
T-CH\cite{Matrix2015} & sym & async & hetero & global &  $P(2P +\lfloor P/2 \rfloor)$ (MRD) &  \\ \hline
DSCR \cite{DSCR2016} & sym & async & hetero & global &  $P(2P +\lfloor P/2 \rfloor)$ (MRD) &  \\ \hline
IDEAL-CH (ours)  & sym & async & hetero & global &    $2{P^\prime}^2$ (MRD) &  \\ \hline
EJS \cite{Lin13}  & sym & async & hetero & global &  $4P(P+1-G)$   &  \\ \hline
FRCH \cite{ChangGY13}  & sym & async & hetero & global &  $N(2N+1)^*$   &  \\ \hline
SARAH \cite{ARCH} & sym & async & hetero & global &    $8N^2+o(N^2)$  &  \\ \hline
SRR \cite{Localimprove2018}  & sym & async & hetero & global &  $2P^2+2P$   &  \\ \hline
ORTHO-CH (ours)  & sym & async & hetero & global &  $2P^2+P$ &    \\ \hline \hline
S-ACH \cite{Bian2013}   & ID & async & hetero & global &    $6L N^2$ (MRD)  &  \\ \hline
E-AHW \cite{Chuang14}   & ID & async & hetero & global &   $(3L+1) NP$ (MRD) &  \\ \hline
CBH \cite{CBH2014}   & ID & async & hetero & local &    $O(L (\max[n_1,n_2])^2)$ (MRD) &  \\ \hline
Adv. rdv-$\eta_1$ \cite{Group2015}   & ID & async & hetero & local &    $(2L+3)n_1 n_2)$ (MRD) & $(2L+3)\frac{n_1 n_2}{G}$ \\ \hline
Two-prime MC \cite{ToN2017}   & ID & async & hetero & local &   $6(\lceil L/4 \rceil *5+6) n_1 n_2$ (MRD) & $\frac{n_1 n_2}{G}+ O((1-\frac{n_1 n_2}{G})^L)$ \\ \hline \hline
QR \cite{Quasi2018}   & sym & async & hetero & global &  $9(\lceil \lceil \log_2 N \rceil /4 \rceil *5+6) n_1 n_2$ &  $\frac{n_1 n_2}{G}+ O((1-\frac{n_1 n_2}{G})^{\log_2 N})$ \\ \hline
Catalan \cite{Chen14}   & sym & async & hetero & global &  $O((\log\log N)  n_1 n_2)$ (MRD) &   \\ \hline
MTP \cite{Improved2015}   & sym & async & hetero & global &  $64 (\lceil \log_2 \log_2 N \rceil+1) (\max[n_1,n_2])^2 $ (MRD) &   \\ \hline
FMR \cite{Chang18}   & sym & async & hetero & global &  $9 (2\lceil \log_2 (\lceil \log_2 N \rceil)\rceil+7) n_1n_2 $ (MRD) &   \\ \hline
QECH \cite{QECH}   & sym & async & hetero & global &  $O((\log N) n_1 n_2)$ &   \\ \hline \hline
AW \cite{AW90}   & sym & sync & homo & local &   & $0.829N$  \\ \hline
random & sym & async & hetero & ind &   & $\frac{n_1 n_2}{G}$   \\
 \hline \hline
\end{tabular}
\label{table:known}
\footnotesize{Remarks: $N$ is the total number of channels, $P$ is a prime not less than $N$, $P^\prime$ is a prime with $P^\prime -2 \sqrt{P^\prime} \ge N$, $n_1$ (resp. $n_2$) is the number of available channels of user 1 (resp. user 2), $G$ is the number of common channels of two users, and $L$ is the length of a user ID (in bits). (MRD) stands for maximum rendezvous diversity. For FRCH, $N \ne ((5+2\alpha)*r-1)/2$ for all integer $\alpha \ge 0$ and odd integer $r \ge 3$. For CACH (resp. FOCAL, SynMAC, M-QCH),  the channel load is $1/u$ (resp. 1, 1, 2/3).
}
\end{center}
\end{table*}
}

\bsec{IDEAL-CH}{ICH}

In this paper, we focus on the sym/async/hetero/global MRP.
As shown in Table \ref{table:known},  CRSEQ \cite{CRSEQ}, JS \cite{JS2011}, DRDS \cite{DRDS13},  T-CH \cite{Matrix2015}, and DSCR \cite{DSCR2016}
are known CH sequences that achieve maximum rendezvous diversity. However, the asymptotic approximation ratio, defined as the ratio of the period to the lower bound $N^2$ when $N \to \infty$, is
still 2.5 for T-CH and DSCR, 3 for CRSEQ and DRDS.
In this section, we tighten the theoretical gap by proposing IDEAL-CH that has the asymptotic approximation ratio 2.

\bsubsec{MACH sequences and matrices}{MACH}

Recall that a periodic CH sequence is said to achieve the maximum rendezvous diversity (MRD) for a CRN with $N$ channels if the two  users rendezvous on every channel  within the period of the sequence. In the following definition, we state formally the mathematical properties for an Asynchronous Channel Hopping sequence with Maximum rendezvous diversity (MACH sequence).

\bdefin{MACH}
An $(N,\peri)$-MACH sequence
$\{c(t), 0 \le t \le p-1\}$ satisfies the following one dimensional maximum rendezvous diversity (1D-MRD) property:
\begin{description}
\item[] {\bf The 1D-MRD property}: for any time shift $0 \le d \le \peri-1$ and any channel $0 \le k \le N-1$, there exists $0 \le t \le \peri-1$ such that
\beq{MACHP1}c(t) =c(t \oplus d)=k,
\eeq
    where $\oplus$ denotes addition modulo $p$.
\end{description}
\edefin

We note that an MACH sequence is simply called a {\em good} sequence in \cite{DRDS13}, and its
connection to the Disjoint Relaxed Difference Set (DRDS) was first made in that paper.
Analogous to the definition of an MACH sequence, we define its 2D version as follows:

\bdefin{MACH2D}
A $\peri \times \peri$ matrix $C=(c_{i,j})$ with $i,j=0,1,\ldots, \peri-1$ is called an $(N,\peri)$-MACH {\em matrix}
if it satisfies the following two-dimensional maximum rendezvous diversity (2D-MRD) property:
\begin{description}
\item[] {\bf The 2D-MRD property}: for any 2D-shift $0 \le \delta, \tau \le \peri-1$ and any channel $0 \le k \le N-1$, there exist $0 \le i, j \le \peri-1$ such that
    \beq{MACHP2} c_{i,j} =c_{i \oplus \delta, j\oplus \tau}=k.
    \eeq
\end{description}
A weaker version of an $(N,\peri)$-MACH matrix is called an $(N,\peri)$-{\em semi}-MACH matrix, in which
the 2D-MRD property may  not be satisfied for $\tau = 0$.
\edefin

Our construction of CH sequences, called the IDEAL-CH, is to construct an $(N,\peri)$-MACH matrix and then use that to construct an
$(N,2\peri^2)$-MACH sequence. In our construction, there are
two elegant mathematical tools for dealing with circular shifts: (i) perfect difference sets \cite{Singer1938} and (ii) ideal matrices \cite{Kumar88}. Intuitively, a perfect  difference set with a period $\peri$ and $k$ elements can be visualized as a dot pattern that has a dot on the 1D-coordinate of an element. Repeat the dot pattern  infinitely often in the line.
Then for any time shift, exactly one pair of dots will overlap in every period of $\peri$.
On the other hand, an ideal matrix can be viewed as a two-dimensional version of a perfect difference set.
A $\peri \times \peri$ ideal matrix has exactly one element in each column and can be visualized as a dot pattern that has a dot on the 2D-coordinate of an element in the matrix. Repeat the dot pattern  infinitely often in the plane. Then except purely vertical shifts,  exactly one pair of dots will overlap within a $\peri \times \peri$ square box for any other two-dimensional shifts  (see Table \ref{table:dot pattern} for an illustration).

Similarly, an $(N,\peri)$-MACH sequence can be repeatedly extended to a periodic sequence in the line. For any time shift, every channel is a rendezvous channel within an interval of length $p$. On the other hand, an $(N,\peri)$-MACH matrix can be repeatedly extended in the plane.
Then for any two-dimensional shift, every channel is a rendezvous channel within a $\peri \times \peri$ square box.

The idea of constructing an $(N,\peri)$-MACH matrix is to {\em first construct a  $\peri \times \peri$ ideal matrix
 and then embed a perfect difference set in each column of that matrix} so that
the overlaps between the constructed matrix and  any two-dimensional circular shift of that matrix contain all the rendezvous channels.
Specifically, we show if $\peri$ is a prime and is equal to $L^2+L+1$ for some prime power $L$, our IDEAL-CH can guarantee $L^2$ rendezvous channels
within the period $2\peri^2$. For IDEAL-CH, the asymptotic approximation ratio is $\frac{2(L^2+L+1)^2}{(L^2)^2}$ and it approaches 2 when $L \to \infty$.

{\tiny
\begin{table}
\begin{center}
\begin{tabular}{||c|c|c|c|c|c|c||c|c|c|c|c|c|c||} \hline
 &  &  & $\bullet$ &  &  &  & &  &  & $\bullet$ &  &  & \\ \hline
 &  &  &  &  &  &   &  &  &  &  &  & & \\ \hline
&  &  &  &  &  &   &  &  &  &  &  & & \\ \hline
 &  & $\bullet$ &  & $\bullet$ &  &  & $\rg \bullet$ &  & $\bullet$ &  & $\bullet$ &  & \\ \hline
&  &  &  &  &  &   &  &  &  &  &  &  & \\ \hline
 & $\bullet$ &  &  &  & $\bullet$ & &  & $\bullet$ &  &  &  & $\bullet$ & \\ \hline
$\bullet$ &  &  &  &  &  & $\rr \bullet$ & $\bullet$ & $\rg \bullet$  &  &  &  &  & $\bullet$ \\ \hline \hline
 &  &  & $\bullet$ &  &  &  & &  &  & $\bullet$ &  &  & \\ \hline
 &  &  &  &  & $\rg \bullet$ &   &  &  & $\rg \bullet$ &  &  & & \\ \hline
&  &  &  & $\rg \bullet$ &  &   &  &  &  & $\rg \bullet$ &  &  & \\ \hline
 &  & $\bullet$ &  & $\bullet$ &  &  &  &  & $\bullet$ &  & $\bullet$ &  & \\ \hline
&  &  &  &  &  &   &  &  &  &  &  & & \\ \hline
 & $\bullet$ &  &  &  & $\bullet$ & &  & $\bullet$ &  &  &  & $\bullet$ & \\ \hline
$\bullet$ &  &  &  &  &  & $\bullet$ & $\bullet$ &  &  &  &  &  & $\bullet$ \\ \hline \hline
\end{tabular}
\caption{A $7 \times 7$ ideal matrix is repeated infinitely often in the plane. It overlaps with the dots of the shifted $7 \times 7$ ideal matrix (marked in green) at exactly one dot pair (the red dot). The 2D shift is represented by $\delta=3$ and $\tau=4$.}
\label{table:dot pattern}
\end{center}
\end{table}
}

\bsubsec{Difference sets}{difference}

In this section, we briefly review the notion of difference sets.

\bdefin{difference}{\bf (Relaxed difference sets (RDS))}
Let $Z_\peri=\{0,1,\ldots, \peri-1\}$.
 A set $D = \{a_0, a_1, \ldots , a_{k-1}\} \subset Z_\peri$  is called a $(\peri,k,\lambda)$-relaxed difference set (RDS) if for every $(\ell\;\mbox{mod}\;\peri) \ne 0$, there exist
at least $\lambda$ ordered pairs $(a_i, a_j)$ such that $a_i-a_j = (\ell\;{\rm mod}\;\peri)$, where $a_i, a_j\in  D$.
A $(\peri,k,1)$-relaxed difference set is said to be {\em perfect} if  there exists
exactly one  ordered pair $(a_i, a_j)$ such that $a_i-a_j = (\ell\;{\rm mod}\;\peri)$ for every $(\ell\;\mbox{mod}\;\peri) \ne 0$.
In this paper, we are only interested in the case $\lambda=1$ and we simply say a set $D$ is a RDS (or a perfect difference set) in $Z_\peri$ when $\lambda=1$.
\edefin

Clearly, if $D = \{a_0, a_1, \ldots , a_{k-1}\}$ is a perfect difference set in $Z_\peri$, then $D_\ell=\{(a_0+\ell)\;\mbox{mod}\;\peri, (a_1+\ell)\;{\rm mod}\;\peri, \ldots , (a_{k-1}+\ell)\;{\rm mod}\;\peri\}$, $\ell=0,1,2,\ldots, \peri-1$, are all perfect difference sets in $Z_\peri$.
Such a rotation property will be used in our embedding of perfect difference sets.
An explicit construction of $(\peri^2+\peri+1, \peri+1,1)$-perfect difference set was shown in \cite{Singer1938} for any $\peri$ that is a prime power. For instance, the set $D=\{0,1,3\}$ is a perfect difference set in $Z_7$.

In view of the mathematical property of a RDS, a periodic CH with $N$ rendezvous channels is equivalent to that there are $N$ {\em disjoint}
RDS in that periodic sequence. Such an equivalent statement was previously made in
 \cite{DRDS13}. Furthermore, the Disjoint Relaxed Difference Set (DRDS) algorithm in \cite{DRDS13} can be used for constructing a CH sequence with maximum rendezvous diversity that has a period of $3N^2$ when the number of channels $N$ is a prime.
 In \cite{Dsetgen2013,Tan2017}, efficient algorithms were proposed to find {\em disjoint} $(\peri^2+\peri+1, \peri+1,1)$-perfect difference sets for a prime power $\peri$. If  the number of disjoint perfect difference sets that can be found for a prime power $\peri$ is not less than the total number of channel $N$,
 then they can be used for constructing CH sequences with maximum rendezvous diversity.
 However, there is no lower bound on the number of disjoint perfect difference sets that can be found for a prime power $\peri$ in \cite{Dsetgen2013,Tan2017}.


\bsubsec{Ideal matrices}{dotmatrix}

In this section, we introduce the notion of an ideal matrix in  \cite{Kumar88}.
As discussed before, an ideal matrix can be viewed as a two-dimensional version of a perfect difference set.

\bdefin{ideal}{\bf (Ideal matrix \cite{Kumar88})}
A binary $(0,1)$ $\peri \times \peri$  matrix $M=(m_{i,j})$ is called an {\em ideal} matrix if it satisfies the following two constraints:
\begin{description}
\item[(i)] Each column of $M$ contains exactly one 1, i.e., for all $j=0,1,2, \ldots, \peri-1$,
\beq{ideal0011}
\sum_{i=0}^{\peri-1}m_{i,j}=1.
\eeq
\item[(ii)] The doubly periodic correlation function $\rho(\cdot,\cdot)$, defined by
\beq{ideal1111}
\rho(\delta, \tau)=\sum_{i=0}^{\peri-1}\sum_{j=0}^{\peri-1} m_{i \oplus \delta, j\oplus \tau}m_{i,j}
\eeq
where $\delta, \tau$ are integers between 0 and $\peri-1$ and $\oplus$ denotes addition modulo $\peri$,
satisfies the condition
\beq{ideal2222}
\rho(\delta, \tau) \le 1
\eeq
whenever either $\delta$ or $\tau$ is nonzero.
\end{description}
\edefin

Since an ideal matrix $M$ contains exactly $\peri$ 1's, we have
\beq{ideal3300}
\rho(0,0)=\peri.
\eeq
On the other hand, we have from \req{ideal0011} that
\bear{ideal3311}
&&\sum_{\delta=0}^{\peri-1} \sum_{\tau=0}^{\peri-1} \rho(\delta, \tau) \nonumber \\
&&=\sum_{i=0}^{\peri-1}\sum_{j=0}^{\peri-1}\sum_{\delta=0}^{\peri-1} \sum_{\tau=0}^{\peri-1}m_{i \oplus \delta, j\oplus \tau}m_{i,j}\nonumber \\
&&=\sum_{j=0}^{\peri-1}\sum_{i=0}^{\peri-1}m_{i,j} \sum_{\tau=0}^{\peri-1}\sum_{\delta=0}^{\peri-1} m_{i \oplus \delta, j\oplus \tau}\nonumber\\
&& =\peri^2 .
\eear
Also, as each column of $M$ contains exactly one 1, we have for $\delta=1,2, \ldots, \peri-1$,
\beq{ideal3322}\rho(\delta, 0)=0.
\eeq
It then follows from \req{ideal2222}, \req{ideal3300}, \req{ideal3311} and \req{ideal3322} that
for $\tau \ne 0$
\beq{ideal3344}
\rho(\delta, \tau)=1 .
\eeq
In view of \req{ideal3322} and \req{ideal3344}, one way to visualize an ideal matrix $M$ as a dot pattern
is to put a dot on the 2D-coordinate of a 1 in $M$. Now repeat the pattern of the matrix infinitely often in the plane. Then
the ideal matrix has the following three important properties:
\begin{description}
\item[(P1)] (No shift) If $(\delta \mod \peri)=(\tau \mod \peri)=0$, all dots overlap.
\item[(P2)] (Purely vertical shifts) For all the purely vertical shifts (along the columns) with $(\tau \mod \peri)=0$ and $(\delta \mod \peri) \ne 0$, no dot will overlap.
 \item[(P3)] (The other shifts) For any the other shifts, i.e., $(\tau \mod \peri)\ne 0$, exactly one pair of dots will overlap.
\end{description}

As each column of an ideal matrix contains exactly one dot, one can view the dot pattern from an ideal matrix as a ``graph'' of a function $f(\cdot)$ with both its domain and range being the set of integers $\{0,1, \ldots, \peri-1\}$. The function $f(\cdot)$ can be characterized as
follows:
\beq{ideal4444}
f(j)=\peri-1-i,
\eeq
where $i$ is uniquely determined by the condition $m_{i,j}=1$.
With such a functional characterization, a $\peri \times \peri$ ideal matrix $M$ can be constructed
when $\peri$ is a prime.

\bthe{ideal}{(\bf The Elliot-Butson construction \cite{RDS1966})}
If $\peri$ is a prime and
\beq{ideal5555}
f(j)=((c_2 j^2 + c_1 j +c_0)\; \mod \; \peri),
\eeq
with
$c_2 \ne 0$,
then the $\peri \times \peri$ matrix $M=(m_{i,j})$ with
\beq{ideal6666}
m_{i,j}=\left \{\begin{array}{ll}
1, &\mbox{if}\; \peri-1-i=f(j), \\
 0, & \mbox{otherwise,}
\end{array}
\right.
\eeq
is an ideal matrix.
\ethe

To see the insight of the Elliot-Butson construction, we note that
$i$ is uniquely determined by $j$ from \req{ideal6666}. Thus,
there is exactly one 1 in each column and \req{ideal0011} is satisfied.
To show  \req{ideal3344}, it suffices to show that
 for any $\tau \ne 0$ and $\delta$  there exists a unique $j$  such that
$m_{i,j}=m_{i \oplus \delta, j \oplus \tau}=1$.
It follows from \req{ideal5555} and \req{ideal6666} that
$$((c_2 j^2 + c_1 j +c_0) \mod \peri)=\peri-1-i$$
and
\bearn
&&((c_2 (j +\tau)^2 +c_1 (j+ \tau) + c_0) \mod \peri )\\
&&=((\peri-1 -i-\delta )\mod \peri ).
\eearn
Solving from these two equations yields
\beq{ideal7777}
(2c_2 \tau j \mod \peri)=((-c_2 \tau^2 -c_1 \tau-\delta)  \mod \peri ).
\eeq
Since $c_2 \ne 0$, $\tau \ne 0$ and $\peri$ is a prime,  there is a unique $j$ satisfying \req{ideal7777}.

One special case of the Elliot-Butson construction  is to choose
\beq{triang1111}
f(j)=\frac{j(j+1)}{2}
\eeq
and this construction is exactly the set of the triangular numbers used in the constructions of the jump columns in CRSEQ \cite{CRSEQ} and T-CH \cite{Matrix2015}. Another example is to choose
\beq{triang1111b}
f(j)=\frac{j(3j-1)}{2}
\eeq
and this construction is exactly the set of the Euler pentagonal numbers used in the constructions of the jump columns in DSCR \cite{DSCR2016}.

\bsubsec{From an ideal matrix to a semi-MACH matrix}{matrix}

To construct a semi-MACH matrix from an ideal matrix, the idea is to replace each column of an ideal matrix by a permutation of $(0,1,2 \ldots, \peri-1)$.
Specifically, define the $i^{th}$-rotation to be the permutation $(i, i\oplus 1, \ldots, i\oplus (\peri-1))$.
Construct a $\peri \times \peri$ matrix $\tilde M =(\tilde m_{i,j})$ by  replacing the $j^{th}$ column of
a $\peri \times \peri$ ideal matrix $M=(m_{i,j})$ by the $(\peri-i)^{th}$-rotation if $m_{i,j}=1$. By doing so, every dot in the ideal matrix is mapped to channel 0 (that serves as an anchor) and every other channel simply rotates around channel 0 in a column.
In the following, we show the conversion for the $7 \times 7$ ideal matrix:
$$
\left [ \begin{array}{lllllll}
0 & 0 & 0 & 1 & 0 & 0 & 0 \\
0 & 0 & 0 & 0 & 0 & 0 & 0 \\
0 & 0 & 0 & 0 & 0 & 0 & 0 \\
0 & 0 & 1 & 0 & 1 & 0 & 0 \\
0 & 0 & 0 & 0 & 0 & 0 & 0 \\
0 & 1 & 0 & 0 & 0 & 1 & 0 \\
1 & 0 & 0 & 0 & 0 & 0 & 1
\end{array}
\right ]
\Rightarrow
\left [ \begin{array}{lllllll}
1 & 2 & 4 & 0 & 4 & 2 & 1 \\
2 & 3 & 5 & 1 & 5 & 3 & 2  \\
3 & 4 & 6 & 2 & 6 & 4 & 3  \\
4 & 5 & 0 & 3 & 0 & 5 & 4 \\
5 & 6 & 1 & 4 & 1 & 6 & 5 \\
6 & 0 & 2 & 5 & 2 & 0 & 6 \\
0 & 1 & 3 & 6 & 3 & 1 & 0 \\
\end{array}
\right ]
$$

One immediate consequence of such a conversion is when one pair of dots overlap, the $\peri$ channels in that column also overlap.
In view of the three properties of an ideal matrix, the matrix $\tilde M$ is a $(\peri,\peri)$-semi-MACH matrix that satisfies the 2D-MRD property except purely vertical shifts, i.e., $\tau=0$ in \req{MACHP2}.

\bsubsec{From  a semi-MACH matrix to an MACH matrix}{matrix2}

To deal with the problem of purely vertical shifts, the idea of T-CH in \cite{Matrix2015} is to concatenate a $\peri \times \peri$ ``stay'' matrix (with all the $\peri$ elements in the $k^{th}$ column being $k$, $k=0,1,2, \ldots, \peri-1$)
and a $\peri \times (\peri+ \lfloor \peri/2 \rfloor)$  ``jump'' matrix with  the $j^{th}$ column taken from the $(j\;\mod\;\peri)^{th}$ column of a semi-MACH matrix.
This results in a $\peri \times (2\peri+ \lfloor \peri/2 \rfloor)$ matrix and thus has a period of $\peri (2\peri+ \lfloor \peri/2 \rfloor)$.
The construction of T-CH shortens the number of ``jump'' columns in CRSEQ \cite{CRSEQ} from $2\peri-1$ to $\peri+ \lfloor \peri/2 \rfloor$. It seems that DSCR \cite{DSCR2016} is somehow equivalent to T-CH. They both are constructed by
concatenating a $\peri \times \peri$ ``stay'' matrix
and a $\peri \times (\peri+ \lfloor \peri/2 \rfloor)$  ``jump'' matrix with  the $j^{th}$ column taken from the $(j\;\mod\;\peri)^{th}$ column of a semi-MACH matrix. The only difference is that they use different quadratic functions in the Elliot-Butson construction for ideal matrices.

Our idea to tackle the problem of purely vertical shifts is to {\em reserve some channels of the $\peri$ channels for embedding relaxed difference sets (RDS)}
that can guarantee  the needed overlaps for purely vertical shifts.

Now we show how to construct an $(L^2,\peri)$-MACH matrix from a $(p,p)$-semi-MACH matrix when $\peri$ is a prime and $\peri$ is equal to $L^2+L+1$ for some  prime power $L$.
The detailed steps are outlined in Algorithm \ref{alg:ICH}.
 Let $D=\{a_0, a_1, \ldots , a_{L}\}$ be an $(L^2+L+1, L+1,1)$-perfect difference set and $\tilde M=(\tilde m_{i,j})$ be
a $(\peri,\peri)$-semi-MACH matrix. Let $D^c= Z_\peri \backslash D=\{b_0, b_1, \ldots, b_{L^2-1}\}$.
The idea is to reserve the $L+1$ channels in $D$ for the perfect difference sets and only use the $L^2$ channels in $D^c$.
The $L+1$ channels in $D$ in the $j^{th}$ column of $\tilde M$ are replaced by channel $(j\;\mod\; L^2)$ and the other $L^2$ channels are
re-mapped to the $L^2$ channels in $\{0,1,2, \ldots, L^2-1\}$.
Specifically, we construct a $\peri \times \peri$ matrix $C=(c_{i,j})$  by the following rule:
\bear{ICH9999}
c_{i,j}= \left\{\begin{array}{ll}
              (j\;\mod\; L^2) &\mbox{if $\tilde m_{i,j} \in D$}\\
              \ell &\mbox{if $\tilde m_{i,j}=b_\ell$}
              \end{array}
              \right..
\eear

For example, the matrix $C$ mapped from the $(7,7)$-semi-MACH matrix
and the perfect difference set $D=\{0,1,3\}$ is shown as follows:
\beq{ideal6699}
\left [ \begin{array}{lllllll}
1 & 2 & 4 & 0 & 4 & 2 & 1 \\
2 & 3 & 5 & 1 & 5 & 3 & 2  \\
3 & 4 & 6 & 2 & 6 & 4 & 3  \\
4 & 5 & 0 & 3 & 0 & 5 & 4 \\
5 & 6 & 1 & 4 & 1 & 6 & 5 \\
6 & 0 & 2 & 5 & 2 & 0 & 6 \\
0 & 1 & 3 & 6 & 3 & 1 & 0 \\
\end{array}
\right ]
\Rightarrow
\left [ \begin{array}{lllllll}
{\bf 0} & 0 & 1 & {\underline {\bf 3}} & 1 & 0 & {\bf 2} \\
0 & {\bf 1} & 2 & {\bf 3} & 2 & {\bf 1} & 0  \\
{\bf 0} & 1 & 3 & 0 & 3 & 1 & {\bf 2}  \\
1 & 2 & {\underline {\bf 2}} & {\bf 3} & {\underline {\bf 0}} & 2 & 1 \\
2 & 3 & {\bf 2} & 1 & {\bf 0} & 3 & 2 \\
3 & {\underline {\bf 1}} & 0 & 2 & 0 & {\underline {\bf 1}} & 3 \\
{\underline {\bf 0}} & {\bf 1} & {\bf 2} & 3 & {\bf 0} & {\bf 1} & {\underline {\bf 2}} \\
\end{array}
\right ]  .
\end{equation}
In this example, the three numbers $0,1,3$ in $D$ of the $j^{th}$ column are mapped to $(j \mod 4)$ for $j=0,1,\ldots, 6$.
Moreover, $D^c=\{2,4,5,6\}$ and these four numbers in the $(7,7)$-semi-MACH matrix  are re-mapped to $\{0,1,2,3\}$, i.e.,
$$2 \mapsto 0,\; 4 \mapsto 1,\; 5 \mapsto 2,\; 6 \mapsto 3.$$
In \req{ideal6699}, we mark the channels that are used for the perfect difference sets in boldface. From the (rotation) property of the perfect difference set, we know for any purely vertical shift, there is an overlap of channel $j$ in column $j$, $j=0,1, \ldots, L^2-1$. Also, those {\em underlined} numbers
are the dots of the $\peri \times \peri$ ideal matrix. These are used as ``anchors'' for any other shifts.

\begin{algorithm}\caption{Construction of an $(L^2,\peri)$-MACH matrix}\label{alg:ICH}

\noindent {\bf Input}  A set of $L^2$ channels $\{0,1,2,\ldots, L^2-1\}$ with $L$ being a prime power  and $L^2+L+1$ being a prime.

\noindent {\bf Output} An $(L^2,\peri)$-MACH matrix $C=(c_{i,j})$ with $p=L^2+L+1$.

\noindent 1: Let $\peri=L^2+L+1$ and construct a $\peri \times \peri$ ideal matrix $M=(m_{i,j})$.

\noindent 2: Construct a $(\peri, \peri)$-semi-MACH matrix $\tilde M=(\tilde m_{i,j})$ by replacing the $j^{th}$  column of
$M$ by the $(\peri-i)^{th}$-rotation of $(0,1,\ldots, \peri-1)$ (for all $j=0,1,\ldots, \peri-1$) if $m_{i,j}=1$.

\noindent 3: Construct a perfect difference set $D=\{a_0, a_1, \ldots , a_{L}\}$ in $Z_{\peri}$.

\noindent 4: Let $D^c= Z_\peri \backslash D=\{b_0, b_1, \ldots, b_{L^2-1}\}$.

\noindent 5: Construct an $(L^2,\peri)$-MACH matrix $C=(c_{i,j})$  by the channel mapping rule in
\req{ICH9999}.


\end{algorithm}

\bthe{ICHmain}
If $L$ is a prime power  and $L^2+L+1$ is a prime, then Algorithm \ref{alg:ICH} constructs an $(L^2,\peri)$-MACH matrix with $p=L^2+L+1$.
\ethe

\bproof
It suffices to prove the 2D-MRD property.
Consider the matrix $C$ from Algorithm \ref{alg:ICH} and the matrix $C^\prime=(c^\prime_{i,j})$ with
$c^\prime_{i,j}=c_{i \oplus \delta, j \oplus \tau}$.
When $\delta=\tau=0$, the two matrices overlap with each other.
For $\delta \ne 0$, we consider the following two cases:

\noindent {\em Case 1}. $\tau=0$:

This corresponds to a purely vertical shift. Since we embed a perfect difference set
 $D$ in the $j^{th}$ column of $C$, the 2D-MRD property is satisfied for channel $j$ in the $j^{th}$ columns of these two matrices, $j=0,1,\ldots, L^2-1$.

\noindent {\em Case 2}. $\tau \ne 0$:

This corresponds to a shift that is not a purely vertical shift. From (P3) of an ideal matrix, there is a column $j_1$ of matrix $C$ that overlaps with a column $j_2$ of matrix $C^\prime$. From the deterministic re-mapping in \req{ICH9999},  the 2D-MRD property is satisfied for all the $L^2$ channels in the overlapped column.
\eproof

\bsubsec{From an MACH matrix to an MACH sequence}{MACHcon}

In this section, we show that one can construct an $(N, 2\peri^2)$-MACH sequence from an $(N,\peri)$-MACH matrix.
The idea to take an $(N,\peri)$-MACH matrix $C=(c_{i,j})$ and concatenate two of them to form a $\peri \times 2\peri$ matrix
$$\tilde C=(\tilde c_{i,j})=(C|C).$$ By doing so, we have $\tilde c_{i,j}=c_{i,(j \mod \peri)}$ for all $i=0,1,\ldots, \peri-1$ and $j=0,1,\ldots, 2\peri-1$.
As the matrix-based construction of CH sequences for T-CH in \cite{Matrix2015},  we then map the matrix $\tilde C=(\tilde c_{i,j})$  to the CH sequence $\{c(t), t=0, 1, \ldots, 2\peri^2-1\}$
by letting $c(t)=\tilde c_{i,j}$ with $i=\lfloor t/(2\peri) \rfloor$ and $j=(t \mod (2\peri))$.
Since $\tilde c_{i,j}=c_{i,(j \mod \peri)}$, this is equivalent to
 letting $c(t)=c_{i,j}$ with $i=\lfloor t/(2\peri) \rfloor$ and $j=(t \mod \peri)$.

For example, concatenating two of the $(4,7)$-MACH matrix in \req{ideal6699}
yields the following $7 \times 14$ matrix:
\beq{ideal6699two}
\begin{array}{cc}
\left [ \begin{array}{lllllll|lllllll}
{\bf 0} & 0 & 1 & {\underline {\bf 3}} & 1 & 0 & {\bf 2} &  {\bf 0} & 0 & 1 & {\underline {\bf 3}} & 1 & 0 & {\bf 2} \\
0 & {\bf 1} & 2 & {\bf 3} & 2 & {\bf 1} & 0 &            0 & {\bf 1} & 2 & {\bf 3} & 2 & {\bf 1} & 0 \\
{\bf 0} & 1 & 3 & 0 & 3 & 1 & {\bf 2} &      {\bf 0} & 1 & 3 & 0 & 3 & 1 & {\bf 2} \\
1 & 2 & {\underline {\bf 2}} & {\bf 3} & {\underline {\bf 0}} & 2 & 1 &    1 & 2 & {\underline {\bf 2}} & {\bf 3} & {\underline {\bf 0}} & 2 & 1 \\
2 & 3 & {\bf 2} & 1 & {\bf 0} & 3 & 2 &        2 & 3 & {\bf 2} & 1 & {\bf 0} & 3 & 2\\
3 & {\underline {\bf 1}} & 0 & 2 & 0 & {\underline {\bf 1}} & 3 &    3 & {\underline {\bf 1}} & 0 & 2 & 0 & {\underline {\bf 1}} & 3  \\
{\underline {\bf 0}} & {\bf 1} & {\bf 2} & 3 & {\bf 0} & {\bf 1} & {\underline {\bf 2}} &  {\underline {\bf 0}} & {\bf 1} & {\bf 2} & 3 & {\bf 0} & {\bf 1} & {\underline {\bf 2}} \\
\end{array}
\right ]  .
\end{array}
\end{equation}
Now the constructed CH sequence with length 98 is then
\bearn
&&0,0,1,3,1,0,2, 0,0,1,3,1,0,2, 0,1,2,3,2,1,0, \\
&& 0,1,2,3,2,1,0, 0,1,3,0,3,1,2,  0,1,3,0,3,1,2,  \\
&& \cdots \\
&& 3,1,0,2,0,1,3, 0,1,2,3,0,1,2,  0,1,2,3,0,1,2.
\eearn

\bthe{MACHcon}
Suppose that the matrix $C=(c_{i,j})$ with $i,j=0,1,\ldots, \peri-1$ is  an $(N,\peri)$-MACH matrix.
Construct the sequence $\{c(t), 0 \le t \le 2\peri^2-1 \}$ by letting
$c(t)=c_{i,j}$ with $i=\lfloor t/(2\peri) \rfloor$ and $j=(t \mod \peri)$.
Then the sequence $\{c(t), 0 \le t \le 2\peri^2-1 \}$ is an $(N, 2\peri^2)$-MACH sequence.
\ethe

\bproof
It suffices to prove the  1D-MRD property for the sequence $\{c(t), 0 \le t \le 2\peri^2-1 \}$, i.e., for any time shift $0 \le d \le 2\peri^2-1$ and any channel $0 \le k \le N-1$, there exists $0 \le t \le 2\peri^2-1$ such that
\beq{MACHP1con}c(t) =c((t + d) \mod (2\peri^2))=k.
\eeq
Let $\delta=\lfloor d/(2\peri) \rfloor$ be the vertical shift  and $\tau=(d \mod (2\peri))$ and be the horizontal shift.
From the matrix-based construction of the CH sequence $\{c(t), 0 \le t \le 2\peri^2-1 \}$,
we can represent such a sequence by the $\peri \times 2\peri$ matrix $\tilde C=(C|C)$.
Similarly, we can also
represent the sequence $\{c((t + d) \mod (2\peri^2)), 0 \le t \le 2\peri^2-1 \}$ by a $\peri \times 2\peri$ matrix $(C_1 | C_2)$ for some $\peri \times \peri$ matrices $C_1$ and $C_2$. 
In view of the 2D-MRD property of the matrix $C=(c_{i,j})$, it suffices to show that either $C_1$ or $C_2$ is
a $\peri \times \peri$ square box in the plane repeated from $C$.

Consider the following two cases:

\noindent {\em Case 1}. $0 \le \tau \le \peri$:

In this case,  the horizontal shift $\tau$ is not greater than $\peri$. Thus, the first matrix $C_1$ is a $\peri \times \peri$ square box in the plane repeated from the matrix $C$. The  2D-MRD property of the matrix $C=(c_{i,j})$  then guarantees the
1D-MRD property of the sequence $\{c((t + d) \mod (2\peri^2)), 0 \le t \le 2\peri^2-1 \}$.
For example, for the CH sequence in \req{ideal6699two}, the sequence $\{c((t + d) \mod 98), 0 \le t \le 97 \}$ in this case can be represented by the matrix $C_1$ marked in {\em red} and the matrix $C_2$ marked in {\em blue}.
\begin{equation*}
\begin{array}{cc}
\left [ \begin{array}{lllllll|lllllll}
{\bf 0} & 0 & 1 & {\underline {\bf 3}} & 1 & 0 & {\bf 2} &  {\bf 0} & 0 & 1 & {\underline {\bf 3}} & 1 & 0 & {\bf 2} \\
0 & {\bf 1} & 2 & {\bf 3} & 2 & {\bf 1} & 0 &            0 & {\bf 1} & 2 & {\bf 3} & 2 & {\bf 1} & 0 \\
{\bf 0} & 1 & 3 & 0 & \rr 3 & \rr 1 & \rr {\bf 2} &   \rr   {\bf 0} & \rr 1 & \rr 3 & \rr 0 & \rb 3 & \rb 1 & \rb   {\bf 2} \\
\rb 1 & \rb 2 & \rb {\underline {\bf 2}} & \rb {\bf 3} & \rr {\underline {\bf 0}} & \rr 2 & \rr 1 &   \rr 1 & \rr 2 & \rr {\underline {\bf 2}} & \rr {\bf 3} & \rb  {\underline {\bf 0}} & \rb 2 & \rb 1 \\
\rb 2 & \rb 3 & \rb {\bf 2} & \rb 1 & \rr {\bf 0} & \rr 3 & \rr 2 &       \rr 2 & \rr 3 & \rr {\bf 2} & \rr  1 & \rb {\bf 0} & \rb 3 & \rb 2\\
\rb 3 & \rb {\underline {\bf 1}} & \rb  0 & \rb 2 & \rr 0 & \rr {\underline {\bf 1}} & \rr 3 &   \rr 3 & \rr {\underline {\bf 1}} &\rr  0 & \rr2 & \rb 0 & \rb {\underline {\bf 1}} & \rb 3  \\
\rb {\underline {\bf 0}} & \rb {\bf 1} & \rb {\bf 2} & \rb 3 & \rr {\bf 0} & \rr {\bf 1} & \rr {\underline {\bf 2}} &  \rr {\underline {\bf 0}} & \rr {\bf 1} & \rr {\bf 2} & \rr 3 & \rb {\bf 0} & \rb {\bf 1} & \rb {\underline {\bf 2}} \\
\rb {\bf 0} & \rb 0 & \rb 1 & \rb {\underline {\bf 3}} & \rr 1 & \rr 0 & \rr {\bf 2} &  \rr {\bf 0} & \rr 0 & \rr 1 & \rr {\underline {\bf 3}} & \rb 1 & \rb 0 & \rb {\bf 2} \\
\rb 0 & \rb {\bf 1} & \rb 2 & \rb {\bf 3} & \rr 2 & \rr {\bf 1} & \rr 0 &           \rr 0 & \rr {\bf 1} & \rr 2 & \rr {\bf 3} &\rb  2 & \rb {\bf 1} & \rb 0 \\
\rb {\bf 0} & \rb 1 & \rb 3 & \rb 0  & 3 & 1 & {\bf 2} &      {\bf 0} & 1 & 3 & 0 & 3 & 1 & {\bf 2} \\
1 & 2 & {\underline {\bf 2}} & {\bf 3} & {\underline {\bf 0}} & 2 & 1 &    1 & 2 & {\underline {\bf 2}} & {\bf 3} & {\underline {\bf 0}} & 2 & 1 \\
2 & 3 & {\bf 2} & 1 & {\bf 0} & 3 & 2 &        2 & 3 & {\bf 2} & 1 & {\bf 0} & 3 & 2\\
3 & {\underline {\bf 1}} & 0 & 2 & 0 & {\underline {\bf 1}} & 3 &    3 & {\underline {\bf 1}} & 0 & 2 & 0 & {\underline {\bf 1}} & 3  \\
{\underline {\bf 0}} & {\bf 1} & {\bf 2} & 3 & {\bf 0} & {\bf 1} & {\underline {\bf 2}} &  {\underline {\bf 0}} & {\bf 1} & {\bf 2} & 3 & {\bf 0} & {\bf 1} & {\underline {\bf 2}} \\
\end{array}
\right ]
\end{array}
\end{equation*}


\noindent {\em Case 2}. $\peri < \tau \le 2\peri-1$:

In this case,  the horizontal shift $\tau$ is larger than $\peri$. Thus, the second matrix $C_2$ is a $\peri \times \peri$ square box in the plane repeated from the matrix $C$. The  2D-MRD property of the matrix $C=(c_{i,j})$  then guarantees the
1D-MRD property of the sequence $\{c((t + d) \mod (2\peri^2)), 0 \le t \le 2\peri^2-1 \}$.
For example, for the CH sequence in \req{ideal6699two}, the sequence $\{c((t+d) \mod 98), 0 \le t \le 97 \}$ in this case can be represented by the matrix $C_1$ marked in {\em red} and the matrix $C_2$ marked in {\em blue}.

\begin{equation*}
\begin{array}{cc}
\left [ \begin{array}{lllllll|lllllll}
{\bf 0} & 0 & 1 & {\underline {\bf 3}} & 1 & 0 & {\bf 2} &  {\bf 0} & 0 & 1 & {\underline {\bf 3}} & 1 & 0 & {\bf 2} \\
0 & {\bf 1} & 2 & {\bf 3} & 2 & {\bf 1} & 0 &            0 & {\bf 1} & 2 & {\bf 3} & 2 & {\bf 1} & 0 \\
{\bf 0} & 1 & 3 & 0 & 3 &  1 & {\bf 2} &      {\bf 0} &  1 &  3 &  0 & \rr 3 & \rr 1 & \rr   {\bf 2} \\
\rr 1 & \rr 2 & \rr {\underline {\bf 2}} & \rr {\bf 3} & \rb {\underline {\bf 0}} & \rb 2 & \rb 1 &   \rb 1 & \rb 2 & \rb {\underline {\bf 2}} & \rb {\bf 3} & \rr  {\underline {\bf 0}} & \rr 2 & \rr 1 \\
\rr 2 & \rr 3 & \rr {\bf 2} & \rr 1 & \rb {\bf 0} & \rb 3 & \rb 2 &       \rb 2 & \rb 3 & \rb {\bf 2} & \rb  1 & \rr {\bf 0} & \rr 3 & \rr 2\\
\rr 3 & \rr {\underline {\bf 1}} & \rr  0 & \rr 2 & \rb 0 & \rb {\underline {\bf 1}} & \rb 3 &   \rb 3 & \rb {\underline {\bf 1}} &\rb  0 & \rb2 & \rr 0 & \rr {\underline {\bf 1}} & \rr 3  \\
\rr {\underline {\bf 0}} & \rr {\bf 1} & \rr {\bf 2} & \rr 3 & \rb {\bf 0} & \rb {\bf 1} & \rb {\underline {\bf 2}} &  \rb {\underline {\bf 0}} & \rb {\bf 1} & \rb {\bf 2} & \rb 3 & \rr {\bf 0} & \rr {\bf 1} & \rr {\underline {\bf 2}} \\
\rr {\bf 0} & \rr 0 & \rr 1 & \rr {\underline {\bf 3}} & \rb 1 & \rb 0 & \rb {\bf 2} &  \rb {\bf 0} & \rb 0 & \rb 1 & \rb {\underline {\bf 3}} & \rr 1 & \rr 0 & \rr {\bf 2} \\
\rr 0 & \rr {\bf 1} & \rr 2 & \rr {\bf 3} & \rb 2 & \rb {\bf 1} & \rb 0 &           \rb 0 & \rb {\bf 1} & \rb 2 & \rb {\bf 3} &\rr  2 & \rr {\bf 1} & \rr 0 \\
\rr {\bf 0} & \rr 1 & \rr 3 & \rr 0  & \rb 3 & \rb 1 & \rb {\bf 2} &     \rb {\bf 0} & \rb 1 & \rb 3 & \rb 0 & 3 & 1 & {\bf 2} \\
1 & 2 & {\underline {\bf 2}} & {\bf 3} & {\underline {\bf 0}} & 2 & 1 &    1 & 2 & {\underline {\bf 2}} & {\bf 3} & {\underline {\bf 0}} & 2 & 1 \\
2 & 3 & {\bf 2} & 1 & {\bf 0} & 3 & 2 &        2 & 3 & {\bf 2} & 1 & {\bf 0} & 3 & 2\\
3 & {\underline {\bf 1}} & 0 & 2 & 0 & {\underline {\bf 1}} & 3 &    3 & {\underline {\bf 1}} & 0 & 2 & 0 & {\underline {\bf 1}} & 3  \\
{\underline {\bf 0}} & {\bf 1} & {\bf 2} & 3 & {\bf 0} & {\bf 1} & {\underline {\bf 2}} &  {\underline {\bf 0}} & {\bf 1} & {\bf 2} & 3 & {\bf 0} & {\bf 1} & {\underline {\bf 2}} \\
\end{array}
\right ]
\end{array}
\end{equation*}
\eproof

With \rthe{MACHcon}, we propose the construction of the IDEAL-CH in Algorithm \ref{alg:ICHcon}.

\begin{algorithm}\caption{The IDEAL-CH}\label{alg:ICHcon}

\noindent {\bf Input}  A set of $L^2$ channels $\{0,1,2,\ldots, L^2-1\}$ with $L$ being a prime power  and $L^2+L+1$ being a prime.

\noindent {\bf Output} A CH sequence $\{c(t), t =0,1,2 \ldots, 2(L^2+L+1)^2-1\}$ with $c(t) \in \{0,1,2,\ldots, L^2-1\}$.

\noindent 1: Use Algorithm \ref{alg:ICH} to construct an $(L^2, \peri)$-MACH matrix with $\peri=L^2+L+1$.

\noindent 2: For $t=0,1,2 \ldots, 2(L^2+L+1)^2-1$, let $c(t)=c_{i,j}$ with $i=\lfloor t/(2\peri) \rfloor$ and $j=(t \mod \peri)$.

\end{algorithm}

As a direct consequence of \rthe{ICHmain} and \rthe{MACHcon}, we have the following corollary.

\bcor{ICH}
If $L$ is a prime power  and $L^2+L+1$ is a prime, then Algorithm \ref{alg:ICHcon} constructs a CH sequence with period $2(L^2+L+1)^2$ that achieves maximum rendezvous diversity for the $L^2$ channels $\{0,1,2,\ldots, L^2-1\}$.
\ecor


\bsubsec{The general construction of an $(N,\peri)$-MACH matrix}{general}

By using a computer search for the set of numbers $L$ with $L$ being a prime power and $L^2+L+1$ being a prime,
 we find $\{2,3,5,8,17,27,41,59,71,89\}$ for $L \le 100$.  There are 4688 positive integers with such properties under 100,000.
For the integers that do not possess such properties, we have to resort to less efficient constructions.
Instead of using a perfect difference set in Step 3 of Algorithm \ref{alg:ICH}, we can use a RDS. It was shown in \cite{LW1997}
that the size of a RDS in $Z_\peri$ is bounded below by $\sqrt{\peri}$.
Here we show how to construct a RDS $D$ in $Z_\peri$ with the size smaller than $2\sqrt{\peri}$.

\index{Relaxed difference set (RDS)}
\index{Difference set!relaxed}
To construct a RDS in $Z_\peri$ for any period $\peri$, the idea is first to place a periodic dot pattern with the period $\Delta$ in the interval $[0,\peri-1]$,
and then add $\Delta$ dots in the interval $[0,\Delta-1]$
as the ``delimiter.'' As there is at least one dot within an interval of length $\Delta$,
the $\Delta$ dots that serve as the delimiter will overlap with at least one dot in any time-shifted dot pattern.
This is stated in the following proposition.

\bprop{RDSp}
For any $\Delta \ge 2$ and $\peri \ge \Delta$,
the set  $D=\{0,1,\ldots, \Delta-1\}\cup\{2\Delta-1, 3\Delta-1, \ldots, \lfloor \peri/\Delta \rfloor \Delta-1\}$ is
a RDS in $Z_\peri$.
\eprop

Such a construction of a RDS can be characterized with two parameters: the period $\peri$ and the spacing $\Delta$.
For example, if we choose $\Delta=5$ for $\peri=23$, then $D=\{0,1,2,3,4,9,14,19\}$ is a RDS in $Z_{23}$ with 8 elements.
Note that the number of elements in $D$ in \rprop{RDSp} is $\Delta+\lfloor \peri/\Delta \rfloor-1$.
To minimize the number of elements in $D$ in \rprop{RDSp}, one may choose the spacing $\Delta=\lceil \sqrt{\peri} \rceil$.
Since $x \le \lceil x \rceil < x+1$ and $\lfloor x \rfloor \le x$, we have
\beq{RDSp1111}
\lceil \sqrt{\peri} \rceil+\lfloor \peri/\lceil \sqrt{\peri} \rceil \rfloor-1 < 2 \sqrt{\peri}.
\eeq
Thus, one can construct a RDS in $Z_\peri$ with the size smaller than $2 \sqrt{\peri}$.

Instead of using a perfect difference set in
Step 3  in Algorithm \ref{alg:ICH}, now we can replace it  by using a RDS in $Z_\peri$ with the spacing $\Delta=\lceil \sqrt{\peri} \rceil$  in \rprop{RDSp}. This leads to the general construction of an $(N,\peri)$-MACH matrix in Algorithm \ref{alg:ICHgen}.

\index{IDEAL-CH!general}
\begin{algorithm}\caption{The general construction of an $(N,\peri)$-MACH matrix}\label{alg:ICHgen}

\noindent {\bf Input}:  A set of $N$ channels $\{0,1,2,\ldots, N-1\}$.

\noindent {\bf Output}: An $(N,p)$-MACH matrix with $\peri$ being the smallest prime such that $\peri-(\lceil \sqrt{\peri} \rceil+\lfloor \peri/\lceil \sqrt{\peri} \rceil \rfloor-1) \ge N$.

\noindent 1: Find the smallest prime $\peri$ such that $\peri-(\lceil \sqrt{\peri} \rceil+\lfloor \peri/\lceil \sqrt{\peri} \rceil \rfloor-1) \ge N$ and construct a $\peri \times \peri$ ideal matrix $M=(m_{i,j})$.

\noindent 2: Construct a $(\peri , \peri)$-semi-MACH matrix $\tilde M=(\tilde m_{i,j})$ by replacing the $j^{th}$  column of
$M$ by the $(\peri-i)^{th}$-rotation of $(0,1,\ldots, \peri-1)$ (for all $j=0,1,\ldots, \peri-1$) if $m_{i,j}=1$.

\noindent 3: Let $\Delta=\lceil \sqrt{\peri} \rceil$. Construct a RDS $D=\{0,1,\ldots, \Delta-1\}\cup\{2\Delta-1, 3\Delta-1, \ldots, \lfloor \peri/\Delta \rfloor \Delta-1\}$ in $Z_{\peri}$.

\noindent 4: Let $D^c= Z_\peri \backslash D=\{b_0, b_1, \ldots, b_{\peri-1-|D|}\}$.

\noindent 5: Construct a $\peri \times \peri$ matrix $C=(c_{i,j})$  by the following channel mapping rule:
\bearn
c_{i,j}= \left\{\begin{array}{ll}
              (j\;\mod\; N) &\mbox{if $\tilde m_{i,j} \in D$}\\
              (\ell\;\mod\; N) &\mbox{if $\tilde m_{i,j}=b_\ell$}
              \end{array}
              \right..
\eearn


\end{algorithm}

\bcor{ICHcor}
The $(N, 2\peri^2)$-MACH sequence constructed by the general construction of an $(N,\peri)$-MACH matrix in Algorithm \ref{alg:ICHgen} and \rthe{MACHcon} has the asymptotic ratio 2.
\ecor

\bproof
Let $D$ be the RDS constructed in \rprop{RDSp} with the spacing $\Delta=\lceil \sqrt{\peri} \rceil$.
Since $|D|\le 2\sqrt{\peri}$,
the number of rendezvous channels $|D^c|= \peri- |D| \ge \peri-2\sqrt{\peri}$.
Thus, the asymptotic  approximation ratio is
\beq{RICH1111}
\frac{2\peri^2}{|D^c|^2}=\frac{2\peri^2}{(\peri-|D|)^2} \to 2,
\eeq
when $\peri \to \infty$.
\eproof

Regarding the computational complexity of Algorithm \ref{alg:ICHgen}, it is clear that Step 2 to Step 5 is $O(\peri^2)$.
Now we show that
the smallest prime $\peri$ with $\peri-(\lceil \sqrt{\peri} \rceil+\lfloor \peri/\lceil \sqrt{\peri} \rceil \rfloor-1) \ge N$
is smaller than $4N$ for $N \ge 16$, and thus the time complexity of Algorithm \ref{alg:ICHgen} is still $O(N^2)$. To see this, we know from the Berstand-Chebyshev Theorem that there exists a prime $\peri^\prime$ with
$2N < \peri^\prime < 4N$. Thus, for $N \ge16$, we have from \req{RDSp1111} that
\bearn
&& \peri^\prime-(\lceil \sqrt{\peri^\prime } \rceil+\lfloor \peri^\prime /\lceil \sqrt{\peri^\prime } \rceil \rfloor-1) \\
&& \ge \peri^\prime -2 \sqrt{\peri^\prime } \\
&&  \ge 2N -2 \sqrt{4N} \\
&& \ge N+(N-4 \sqrt{N}) \ge N.
\eearn

\bsec{ORTHO-CH}{orthoCH}

To use the IDEAL-CH in the sym/async/hetero/global MRP, each user can simply replace at random those channels not in its available channel set by some channels in its available channel set. By doing so, the two users are still guaranteed to rendezvous on {\em every} commonly available channel in the period of the IDEAL-CH sequence.
Thus, the MCTTR is bounded by the period of the IDEAL-CH sequence.

In this section, we consider a weaker requirement that only needs the two users to rendezvous on {\em one} commonly available channel in a period.
For this, we propose a channel hopping sequence, called ORTHO-CH, that can guarantee  the rendezvous of the two users within a period of the ORTHO-CH sequence.
When the available channels of a user  is a subset of $\{0,1, \ldots, N-1\}$, the period of our ORTHO-CH sequence is $(2\peri +1)\peri$,
where $\peri$ is the smallest prime not less than $N$. Thus, ORTHO-CH has the MTTR bound $(2\peri +1)\peri$.

\bsubsec{Orthogonal MACH matrices}{ortho}

For the construction of the ORTHO-CH sequence,
we introduce a new notion of {\em orthogonal MACH matrices}.

\bdefin{MACH2Dortho}
A set of $\peri \times \peri$ matrices $\{C^{(r)}=(c^{(r)}_{i,j}), r=1,2, \ldots, K\}$ is called a set of {\em orthogonal} $(N,\peri)$-MACH {\em matrices}
if it satisfies the following two properties:
\begin{description}
\item[(i)] {\bf The cover property}: for any channel $0 \le k \le N-1$, it appears at least once in every column of every matrix in  the set of matrices.
\item[(ii)] {\bf The 2D-MRD property}: for any two different matrices $r_1$ and $r_2$,  any 2D-shift $0 \le \delta, \tau \le \peri-1$, and any channel $0 \le k \le N-1$, there exist $0 \le i, j \le \peri-1$ such that
    \beq{MACHP2ortho} c^{(r_1)}_{i,j} =c^{(r_2)}_{i \oplus \delta, j\oplus \tau}=k.
    \eeq
\end{description}
\edefin

We note that the cover property is not needed in \rdef{MACH2D} for an $(N, \peri)$-MACH matrix even though such a property is satisfied in our constructions of the $(N, \peri)$-MACH matrices in Algorithm \ref{alg:ICH} and Algorithm \ref{alg:ICHgen}. Intuitively, one can view an
$(N, \peri)$-MACH matrix as a matrix that is ``orthogonal'' to itself in the sense of the 2D-MRD property.

We choose the phrase ``orthogonal'' from the notion of orthogonal Latin squares \cite{Euler}. In a $\peri \times \peri$ Latin square, every row and every column is a permutation of $\{0,1,\ldots,\peri-1\}$. Two $\peri \times \peri$ Latin squares $C^{(r_1)}=(c^{(r_1)}_{i,j})$ and $C^{(r_2)}=(c^{(r_2)}_{i,j})$ are said to be orthogonal if the $\peri^2$ ordered pairs $(c^{(r_1)}_{i,j}, c^{(r_2)}_{i,j})$, $i,j=0,1,\ldots, \peri-1$ are all different. The number of mutually orthogonal $\peri \times \peri$ Latin squares is bounded by $\peri-1$ and it is achieved when $\peri$ is a prime power. In particular,
when $\peri$ is a prime, the $\peri-1$ orthogonal Latin squares can be constructed by
letting $c^{(r)}_{i,j}=(r \cdot i +j) \mod \peri$, $r=1, 2, \ldots, \peri-1$. In the following theorem, we show such a construction  also leads to
a set of $\peri-1$  orthogonal $(\peri,\peri)$-MACH matrices.

\bthe{ortho}
Suppose that $\peri$ is a prime. For $r=1,2, \ldots, \peri-1$, $0 \le i, j \le \peri-1$,
let
\beq{ortho1111}
c^{(r)}_{i,j}=(r \cdot i +j) \mod \peri.
\eeq
Then the set of matrices $\{C^{(r)}=(c^{(r)}_{i,j}), r=1,2, \ldots, \peri-1\}$ is a set of orthogonal $(\peri,\peri)$-MACH matrices.
\ethe

\bproof
As $r \ne 0$, we have from \req{ortho1111}, every channel $0 \le k \le \peri-1$ appears exactly once in every column of every matrix in the set, and thus the cover property is satisfied.
To show the 2D-MRD property, for $r_1 \ne r_2$, any 2D-shift $0 \le \delta, \tau \le \peri-1$ and any channel $0 \le k \le N-1$, we let $i^*$ be the unique solution of the following equation:
\beq{ortho2222}
((r_1-r_2) \cdot i \mod \peri)=((r_2 \cdot \delta +\tau) \mod \peri),
\eeq
and
\beq{ortho3333}
j^*=((k- r_1 \cdot i^*) \mod \peri).
\eeq
Then we have from \req{ortho3333} and \req{ortho1111} that
$$c^{(r_1)}_{i^*,j^*}=k.$$
Also, it is easy to see from \req{ortho1111} and \req{ortho2222} that
\bearn
&& c^{(r_2)}_{i^*\oplus\delta,j^*\oplus\tau}\\
&&=c^{(r_2)}_{{(i^*+\delta) \mod \peri},{(j^*+\tau) \mod \peri}}\\
&&=(r_2 \cdot (i^*+\delta) +(j^*+\tau)) \mod \peri \\
&&=( r_2 \cdot i^* +(r_2 \cdot \delta +\tau) + j^* ) \mod \peri \\
&&=(r_1 \cdot i^* + j^*)  \mod \peri \\
&&= c^{(r_1)}_{i^*, j^*}.
\eearn
\eproof

For $\peri=5$, the four $(5,5)$-MACH matrices are as follows:
\beq{ortho5555}
\begin{array}{cc}
\left [ \begin{array}{lllll}
0& 1 & 2 & 3 & 4 \\
1& 2 & 3 & 4 & 0 \\
2& 3 & 4 & 0 & 1 \\
3& 4 & 0 & 1 & 2 \\
4& 0 & 1 & 2 & 3
\end{array}
\right ] , &
\left [ \begin{array}{lllll}
0& 1 & 2 & 3 & 4 \\
2& 3 & 4 & 0 & 1 \\
4& 0 & 1 & 2 & 3 \\
1& 2 & 3 & 4 & 0 \\
3& 4 & 0 & 1 & 2
\end{array}
\right ] ,
\end{array}
\end{equation}
\beq{ortho5555b}
\begin{array}{cc}
\left [ \begin{array}{lllll}
0& 1 & 2 & 3 & 4 \\
3& 4 & 0 & 1 & 2 \\
1& 2 & 3 & 4 & 0 \\
4& 0 & 1 & 2 & 3 \\
2& 3 & 4 & 0 & 1
\end{array}
\right ] , &
\left [ \begin{array}{lllll}
0& 1 & 2 & 3 & 4 \\
4& 0 & 1 & 2 & 3 \\
3& 4 & 0 & 1 & 2 \\
2& 3 & 4 & 0 & 1 \\
1& 2 & 3 & 4 & 0
\end{array}
\right ] .
\end{array}
\end{equation}
These four matrices are also mutually orthogonal Latin squares.

\bsubsec{From orthogonal MACH matrices to asynchronous CH sequences}{orthocon}

In this section, we show that one can construct the ORTHO-CH sequence from a set of orthogonal $(\peri, \peri)$-MACH matrices, $\{C^{(r)}, r=1,2, \ldots, \peri-1\}$.
The idea is quite similar to the quasi-random algorithm in \cite{Quasi2018}. Each user selects a nonzero channel $r$ from its available channel set  as its ID channel. Then construct the $\peri \times (2\peri+1)$ matrix ${\tilde C}^{(r)}$ by concatenating the ID column ${\bf r}$ (that stays on channel $r$) and two identical matrices of $C^{(r)}$, i.e.,
\beq{ortho6666}
{\tilde C}^{(r)}=({\tilde c}^{(r)}_{i,j})=({\bf r}|C^{(r)}|C^{(r)}).
\eeq
As in the matrix-based construction for IDEAL-CH, we then map the $\peri \times (2\peri+1)$ matrix $\tilde C^{(r)}$ to
the periodic ORTHO-CH sequence with period $(2\peri+1)\peri$. Channels that are not in the available channel set are randomly replaced by channels in the available channel set.
If the two users select the same ID channel, then both users are guaranteed to rendezvous from the cover property of  a set of orthogonal MACH matrices. On the other hand, if the two users select two different ID channels, then both users are guaranteed to rendezvous on every commonly available channel from the 2D-MRD property of  a set of orthogonal MACH matrices. As such, the two users are guaranteed to rendezvous within the period $(2\peri+1)\peri$.
 The detailed construction is shown in Algorithm \ref{alg:ORTHOcon}.

\begin{algorithm}\caption{The ORTHO-CH}\label{alg:ORTHOcon}

\noindent {\bf Input}  A set of available channels ${\bf c}$ that is a subset of $\{0,1, \ldots, N-1\}$.

\noindent {\bf Output} A CH sequence $\{c(t), t =0,1,2 \ldots, \peri (2\peri+1)-1\}$ with $c(t) \in {\bf c}$, where $\peri$ is the smallest prime not less than $N$.

\noindent 1: If channel 0 is the only channel in ${\bf c}$, output $c(t)=0$ for all $t =0,1,2 \ldots, \peri (2\peri+1)-1$.

\noindent 2: Randomly select a nonzero channel $r$ from ${\bf c}$ as the ID channel.

\noindent 3: Find the smallest prime $\peri$ such that $\peri \ge N$ and construct a $\peri \times \peri$ matrix $C^{(r)}=(c^{(r)}_{i,j})$ by letting
$$c^{(r)}_{i,j}=(r \cdot i +j) \mod \peri.$$

\noindent 4: Let ${\bf r}$ be the $\peri \times 1$ column vector with all its elements being $r$.
Construct the $\peri \times (2\peri+1)$ matrix ${\tilde C}^{(r)}$ by concatenating the column vector ${\bf r}$ and two identical matrices of $C^{(r)}$, i.e.,
$${\tilde C}^{(r)}=({\tilde c}^{(r)}_{i,j})=({\bf r}|C^{(r)}|C^{(r)}).$$

\noindent 5: For $t=0,1,2 \ldots, \peri (2\peri+1)-1$, let $c(t)={\tilde c}^{(r)}_{i,j}$ with $i=\lfloor t/(2\peri+1) \rfloor$ and $j=(t \mod (2\peri+1))$.

\noindent 6: If $c(t)$ is not in ${\bf c}$, replace it at random by a channel in ${\bf c}$.

\end{algorithm}

For example, if $N=4$, then $\peri=5$. Suppose that  the available channel set ${\bf c}=\{0,1,3\}$ and channel 3 is selected as the ID channel.
From \req{ortho5555b}, the $5 \times 11$ matrix $\tilde C$ is constructed as follows:
\beq{ortho5577}
\left [ \begin{array}{l|lllll|lllll}
3 & 0& 1 & 2 & 3 & 4 & 0& 1 & 2 & 3 & 4 \\
3 & 3& 4 & 0 & 1 & 2　&3& 4 & 0 & 1 & 2 \\
3 & 1& 2 & 3 & 4 & 0 & 1& 2 & 3 & 4 & 0 \\
3 & 4& 0 & 1 & 2 & 3 & 4& 0 & 1 & 2 & 3\\
3 & 2& 3 & 4 & 0 & 1 & 2& 3 & 4 & 0 & 1
\end{array}
\right ].
\eeq
Now replace the channels $2$ and $4$ by randomly selected channels  in ${\bf c}$ (marked in R) leads to the following CH sequence:
\bearn
&& 3 , 0, 1, R, 3, R, 0, 1, R, 3, R, \\
&& 3 , 3, R, 0, 1, R, 3, R, 0, 1, R, \\
&& 3 , 1, R , 3 , R , 0 , 1, R , 3 , R , 0 ,\\
&& 3 , R, 0 , 1 , R , 3 , R, 0 , 1 , R , 3 ,\\
&&3 , R, 3 , R , 0 , 1 , R, 3 , R , 0 , 1 .
\eearn

\bthe{orthomain}
Suppose that  user $i$, $i=1$ and 2, have the available channel sets ${\bf c}_i$, $i=1$ and 2, that are subsets of $\{0,1,2 \ldots, N-1\}$ and that both users use the
ORTHO-CH in Algorithm \ref{alg:ORTHOcon} to generate its CH sequence.
If there is at least one commonly available channel, i.e., ${\bf c}_1 \cap {\bf c}_2 \ne \varnothing$,
then both users are guaranteed to  rendezvous within $(2 \peri+1)\peri$ time slots for any clock drift $d$ between these two users,
where $\peri$ is the smallest prime not less than $N$.
\ethe

\bproof
The case that one of the two users only has channel 0 in its available channel set is trivial as that user will stay on channel 0 all the time.
Thus, it suffices to consider the case that both users have at least one  channel that is not channel 0. Let $r_i$ be the ID channel selected by user $k$, $k=1$ and 2, and $\{c_k(t), t=0,1,\ldots \}$ be the CH sequence of user $k$ from the ORTHO-CH in Algorithm \ref{alg:ORTHOcon}.
Under the assumption that there is at least one commonly available channel, we need to show that there exists $0 \le t \le (2\peri+1)\peri-1$
such that
for any time shift $0 \le d \le (2\peri+1)\peri-1$,
\beq{ortho7777}c_1(t)=c_2(t+d).
\eeq
Let $\delta=\lfloor d/(2\peri+1) \rfloor$ be the vertical shift  and $\tau=(d \mod (2\peri+1))$ be the horizontal shift.
In view of the matrix-based construction of CH sequences, the condition in \req{ortho7777}
holds if and only if for any $0 \le \delta \le \peri-1$ and $0 \le \tau \le 2\peri-1$,  there exist $0 \le i \le \peri-1$ and $0 \le j \le 2\peri-1$ such that
    \beq{ortho8888} {\tilde c}^{(r_1)}_{i,j} ={\tilde c}^{(r_2)}_{i \oplus \delta, j\oplus \tau},
    \eeq
where ${\tilde C}^{(r_k)}=({\tilde c}^{(r_k)}_{i,j})$, $k=1$ and 2, are the $\peri \times (2\peri+1)$ matrices defined in \req{ortho6666}.
Now consider the following two cases:

\noindent {\em Case 1}. $r_1=r_2$:

In this case, both users select the same ID channel from their available channel sets. As such, $r_1$ is in the available channel set of user 2.
If $\tau=0$, then both users rendezvous on the same ID channel of column 0, i.e., for all $0 \le i \le \peri-1$,
 \beq{ortho88881} {\tilde c}^{(r_1)}_{i,0} ={\tilde c}^{(r_2)}_{i \oplus \delta,  0}=r_1=r_2.
    \eeq
 On the other hand, if $\tau \ne 0$,
it then follows from the cover property that
 column $\tau$ of ${\tilde C}^{(r_2)}$ contains at least one $r_1$. Thus, there exists $0 \le i^* \le \peri-1$ such that the $(i^* \oplus \delta)^{th}$ element of column $\tau$ of ${\tilde C}^{(r_2)}$ is $r_1$, i.e., ${\tilde c}^{(r_2)}_{i^* \oplus \delta,  \tau}=r_1$.
 Since the $\peri$ elements in column 0 of ${\tilde C}^{(r_1)}$ are all $r_1$, we then have
    \beq{ortho88882} {\tilde c}^{(r_1)}_{i^*,0} ={\tilde c}^{(r_2)}_{i^* \oplus \delta,  \tau}=r_1.
    \eeq
For example, suppose that  $\{c_1(t), 0 \le t \le 54 \}$ is the CH sequence in \req{ortho5577}. Then the sequence $\{c_2((t + d) \mod 55), 0 \le t \le 54 \}$ in this case can be represented by the concatenation of its ID column, the  first $C^{(r_2)}$ matrix,  and the second $C^{(r_2)}$ matrix.
The overlaps of the ID column (resp. the first matrix, the second matrix) with the sequence $\{c_1(t), 0 \le t \le 54 \}$ is marked in {\em green} (resp. {\em red}, {\em blue}) in \req{ortho5588}. Note that the 5 elements marked in green forms a permutation of $\{0,1,2,3,4\}$.

\beq{ortho5588}
\left [ \begin{array}{l|lllll|lllll}
3 & 0& 1 & 2 & 3 & 4 & 0& 1 & 2 & 3 & 4 \\
3 & 3& 4 & 0 & 1 & 2　&3& 4 & 0 & 1 & 2 \\
3 & 1& 2 & \rg 3 & \rr 4 & \rr 0 & \rr 1& \rr 2 & \rr 3 & \rb 4 & \rb 0 \\
\rb 3 & \rb 4& \rb 0 & \rg 1 & \rr 2 & \rr 3 & \rr 4& \rr 0 & \rr 1 & \rb 2 & \rb 3\\
\rb 3 & \rb 2& \rb 3 & \rg 4 & \rr 0 & \rr 1 & \rr 2& \rr 3 & \rr 4 & \rb 0 & \rb 1 \\
\rb 3 & \rb 0& \rb 1 & \rg 2 & \rr 3 & \rr 4 & \rr 0& \rr 1 & \rr 2 & \rb 3 & \rb 4 \\
\rb 3 & \rb 3& \rb 4 & \rg 0 & \rr 1 & \rr 2 & \rr 3& \rr 4 & \rr 0 & \rb 1 & \rb 2 \\
\rb 3 & \rb 1& \rb 2 & 3 & 4 & 0 & 1& 2 & 3 & 4 & 0 \\
3 & 4& 0 & 1 & 2 & 3 & 4& 0 & 1 & 2 & 3\\
3 & 2& 3 & 4 & 0 & 1 & 2& 3 & 4 & 0 & 1
\end{array}
\right ].
\eeq

\noindent {\em Case 2}. $r_1 \ne r_2$:

In this case, the two users select two different ID channels. As such, their CH sequences are constructed from two mutually orthogonal MACH matrices.
As in the proof of \rthe{MACHcon}, we consider the following two subcases:

\noindent {\em Case 2.1}. $0 \le \tau \le \peri$:

In this subcase, the  {\em first} matrix $C^{(r_2)}$ of ${\tilde C}^{(r_2)}$ overlaps with a $\peri \times \peri$ square box in the plane repeated from $C^{(r_1)}$ (see, e.g., the square marked in red in \req{ortho5588}).
Under the assumption that there is at least one commonly available channel, the condition in \req{ortho8888} follows immediately from
the 2D-MRD property of two mutually orthogonal MACH matrices.

\noindent {\em Case 2.2}. $\peri < \tau \le 2\peri-1$:

In this subcase, the  {\em second} matrix $C^{(r_2)}$ of ${\tilde C}^{(r_2)}$ overlaps with a $\peri \times \peri$ square box in the plane repeated from $C^{(r_1)}$. Once again,
under the assumption that there is at least one commonly available channel, the condition in \req{ortho8888} follows immediately from
the 2D-MRD property of two mutually orthogonal MACH matrices.

\eproof

In comparison with FRCH in \cite{ChangGY13},  ORTHO-CH has the same MTTR bound $(2N+1)N$ if $N$ is a prime.
Note that $N \ne ((5+2\alpha)*r-1)/2$  is required for FRCH. For instance, if $\alpha=0$ and $r=3$, then FRCH does not guarantee the rendezvous of the two users for $N=7$.
To achieve such an MTTR bound, FRCH has to remap the channels that are not in the available channel set according to  a specific remapping rule. For ORTHO-CH, such replacements can be chosen randomly.
In comparison with the Sequence-Rotating-Rendezvous (SRR) algorithm in \cite{Localimprove2018}, our ORTHO-CH sequence reduces the MTTR from $2\peri^2 + 2\peri$ to $(2\peri+1)\peri$.
Both constructions are similar in the sense that they both are based on the two mathematical properties  of orthogonal MACH matrices (and thus the proofs are also similar). The key difference is that the ORTHO-CH sequence is periodic while
the SRR sequence is not. In practice, there might be  a nonzero probability that the two users may not rendezvous even when they both hop on a common channel. In such a setting,  there might be a problem for the SRR algorithm when the two users select the same ID channel and they miss their rendezvous on the ID channel. In that sense,
ORTHO-CH is more robust than  SRR.  Similarly, IDEAL-CH is more robust than ORTH-TH as every commonly available channel is a rendezvous channel in IDEAL-CH. However, the period $\peri$ of the general IDEAL-CH is the smallest prime with
$\peri-(\lceil \sqrt{\peri} \rceil+\lfloor \peri/\lceil \sqrt{\peri} \rceil \rfloor-1) \ge N$, which is in general larger than the period of ORTHO-CH for the same  total number of channels $N$. 

As described in the book \cite{Book}, both IDEAL-CH and ORTHO-CH sequences are known as {\em global} sequences as 
they are constructed from all the $N$ channel and then replace those channels not in the available channel set of a user by some channels in its available channel set. Another approach is to construct CH sequences
directly from the available channel sets of users.   Such sequences are called {\em local} sequences, e.g., QR \cite{Quasi2018}, Catalan \cite{Chen14}, MTP \cite{Improved2015}, FMR \cite{Chang18}, and QECH \cite{QECH}. When the numbers of channels of the two users, $n_1$ and $n_2$ are $O(N^\alpha)$ for some $0 < \alpha <1$, then the MTTR bounds from these local sequences are $o(N^2)$ (see Table \ref{table:known}) and thus better than those from global sequences.
On the other hand, if $n_1$ and $n_2$ are linear in $N$, then the $O(N^2)$  of MTTR bounds of global sequences are better than those of local sequences.

\bsec{Conclusion}{con}

By embedding difference sets into an ideal matrix, we are able to tighten the theoretical gap of the asymptotic approximation ratio
for CH sequences with maximum rendezvous diversity from 2.5 to 2. It seems  difficult to further reduce the ratio.
This is mainly due to the  nature of asynchronous clocks of the two users. As in the case that the slot boundaries are not aligned,
one needs to increase the slot size by a factor of 2 to ensure a sufficient overlap of time for rendezvous.
Finally, we conclude the paper by quoting the following comment from the end of the excellent book \cite{Book}"

\begin{center}
{\em ``First of all, closing the gap between the lower bounds on the maximum time to
rendezvous in worst-case situations and the upper bounds by the presented algorithms
will likely be a long term project.''}
\end{center}

\end{document}